\documentclass[twocolumn,showpacs,superscriptaddress,reprint,longbibliography]{revtex4-1}
\usepackage{amsmath, amsfonts, amssymb}
\usepackage{enumerate}
\usepackage{graphicx}
\usepackage{color}
\usepackage{float}
\usepackage[breaklinks]{hyperref}
\usepackage[hyphenbreaks]{breakurl}
\usepackage{paralist}
\usepackage{multirow}
\usepackage{diagbox}
\usepackage{makecell}

\begin{document}


\title{A Green's function approach to Topological Insulator junctions with magnetic and superconducting regions}

\newcommand{\aalto}{Department of Applied Physics, Aalto University, 00076 Aalto, Finland}
\newcommand{\bogota}{Departamento de F\'{\i}sica,
	Universidad Nacional de Colombia, Bogot\'a, Colombia}
\newcommand{\bosque}{Departamento de F\'{\i}sica,
	Universidad el Bosque, Bogot\'a, Colombia}
\newcommand{\uam}{Departamento de F\'{\i}sica Te\'orica de la Materia Condensada, Condensed Matter Physics Center (IFIMAC) and Instituto Nicol\'as Cabrera, Universidad Aut\'onoma de Madrid, Spain}
\newcommand{\order}{Preliminary author position}

\author{Oscar E. Casas}
\affiliation{\bogota}

\author{Shirley G\'omez P\'aez}
\affiliation{\bogota}
\affiliation{\bosque}

\author{William J. Herrera}
\affiliation{\bogota}

\date{\today}


\begin{abstract}

This work presents a Green's function approach, originally implemented in graphene with well-defined edges, to the surface of a strong 3D Topological Insulator (TI) with a sequence of proximitized superconducting (S) and ferromagnetic (F) surfaces. This consists of the derivation of the Green's functions for each region by the asymptotic solutions method, and their coupling by a tight-binding Hamiltonian with the Dyson equation to obtain the full Green's functions of the system. These functions allow the direct calculation of the momentum-resolved spectral density of states, the identification of subgap interface states, and the derivation of the differential conductance for a wide variety of configurations of the junctions. We illustrate the application of this method for some simple systems with two and three regions, finding the characteristic chiral state of the Quantum Anomalous Hall Effect (QAHE) at the NF interfaces, and chiral Majorana modes at the NS interfaces. Finally, we discuss some geometrical effects present in three-region junctions such as weak Fabry-P\'erot resonances and Andreev bound states.
\end{abstract}

\maketitle


\section{Introduction}

The unusual electronic properties of Topological Insulators and their underlying physics have been subject of intense research in the last decade \cite{Zhang_2011,Kane_2010,Ando_2013,Bernevig_book}. The relativistic and helical nature of the surface states, combined with topological protection against time-reversal symmetry perturbations, make these materials suitable candidates for the construction of nanodevices free of dissipation and decoherence \cite{Yokoyama,Zhang_2011,Kane_2010}. Among these materials stand out the family of strong TIs Bi$_{2}$Se$_{3}$, Bi$_{2}$Te$_{3}$, and Sb$_{2}$Te$_{3}$ whose surface states present a conical dispersion relation and left-handed helical spin texture, that can be described by a simple relativistic model at the $\Gamma$ point in momentum space \cite{Zhang_2011,Zhang_2009,Zhang_2010,Silvestrov_2012}. These particular features have already been observed by spin-resolved ARPES \cite{Xia_2009,Chen_2009}, and in some transport experiments at very low temperatures, either by the weak anti-localization \cite{Chen_2010_WAL,Chen_2011,Steinberg_2011,Seung_2011,Cha_2012,Wang_2018} or Shubnikov-de Haas oscillations analysis \cite{Xiong_2012,Shrestha_2014,Vries_2017,Akiyama_2018}. 

The introduction of the ferromagnetic and superconducting order on these materials surface by proximity effect, also give rise to a rich and novel phenomenology. Magnetization induced by a ferromagnetic insulator results in the QAHE, which exhibits a chiral bound state at the system's boundaries \cite{Zhang_2008, Nagaosa_2010_F, Chang_2013, Xu_2014, Brey_2014, Yoshimi_2015,Kubota_2016}. On the other hand, the proximity effect with conventional superconductor results in an effective topological spinless $p$-wave order parameter. Its zero energy Andreev bound states at vortices and FS interfaces constitute Majorana modes, that could be implemented in topological quantum computations technologies \cite{Kane_2008, Xu_2015,Sun_2016_TI, Sun_2017}. The robust topological character of these interface states lies in the bulk-boundary correspondence, which relates the change of some topological invariant between two phases with the emergence of interface bound states \cite{Zhang_2011,Kane_2010,Ando_2013,Bernevig_book}. Additionally, in these systems the NS interfaces would present specular Andreev reflections for low doping of the N region analogous to those predicted for graphene, provided that transport is restricted to the surface \cite{Beenakker_2006, Nagaosa_2010, Procolo_2018}.

The study of the electrical transport properties of junctions on the TI's surface with magnetic, superconducting or mixed regions requires the explicit calculation of the differential conductance and the identification of the transport channels involved. In a first approximation, this problem has been addressed through the scattering matrix or BTK formalism, where the scattering solutions and the associated reflection and transmission coefficients determine the conductance of the system. Besides, this approach considers the different dispersion processes at the NS interface, making it physically intuitive and computationally undemanding for simple junctions \cite{Kane_2009,Tanaka_2009,Nagaosa_2010,Linder_2010,Mondal_2010,Soodchomshom_2010,Yokoyama_2012,Lababidi_2012,Snelder_2013,Cheng_2014,Suwanvarangkoon_2011,Vali_2012,Vali_2012_CM,Niu_2010,Tkachov_2013,Yang_2014,Vali_2014_0,Vali_2014,Cheng_2018}. Other works use sophisticated and exhaustive Green's function techniques that allow the direct calculation of all transport observables and can be implemented even for systems with time-dependent perturbations, particle interactions and disorder \cite{Li_2014,Lu_2015,Snelder_2015,Burset_2015, Acero_2015, Zyuzin_2016, Mohammad_2016,Mohammad_2017,Lu_2018}. The study of the transport properties of heterostructures using the Green's function formalism usually requires a significant amount of numerical calculation, except in some simple non-interacting systems in the stationary regime. 

For translationally-invariant 2D systems, the McMillan's Green's functions method has the advantage of combining the rigor and generality of Green's functions approaches with the simplicity and physical intuition of the BTK formalism, since Green's functions are calculated analytically from the full wave function of the system, that is, from the linear combination  of the scattering states present in all the subregions of the junction \cite{McMillan_1968,Lu_2015,Burset_2015,Lu_2018}.
However, the Green's functions obtained by this method are exclusive of each system and cannot be implemented to describe other configurations. Thus, this method becomes inconvenient for the study of junctions with many subregions, as in the case of the scattering matrix formalism.    

In this paper we study the transport properties of junctions on the surface of Bi$_{2}$Se$_{3}$ in contact with ferromagnetic and conventional superconductors materials. For this, we adapted a Green's functions approach that had previously been applied in graphene with well-defined edges \cite{Herrera_2010,Gomez_2011,Casas_2019_1,Casas_2019}. In this approach, the Green's functions of N, F, or S regions in a junction are calculated from the asymptotic solutions method. Then, they are coupled by a tight-binding Hamiltonian with the Dyson equation, obtaining the Green's functions of the complete system. This allows the study of a wide variety of junctions just by coupling the same basic elements in different sequences. Thus, the analytical Green's functions obtained allow the direct calculation of the momentum-resolved spectral density, the DOS, and the differential conductance within the framework of the Hamiltonian approach \cite{Yeyati_1996}. In some simple cases, it is possible to derive the dispersion relations of interface bound states through the analysis of the Green's functions poles. Furthermore, our approach can be implemented for the study of junctions with an infinite number of regions like superlattices without considerable computational cost \cite{Camilo_2019}. 

The method is illustrated by application to some simple junctions studied in the literature. The Ferromagnetic/Ferromagnetic (FF) junction exhibits chiral interface states associated with the QAHE, and the Ferromagnetic/Superconductor (FS) junction presents chiral Majorana Interface Bound States (IBS), which is in complete agreement with the results reported in the literature \cite{Zhang_2008,Nagaosa_2010_F,Brey_2014,Kane_2008,Tanaka_2009,Nagaosa_2010}. Besides, we present some innovative results for the transport properties of junctions with three regions such as NFN, NFS, FNF, FNS where some geometric resonances associated with quasi-bound states were observed, due to the finite size of the central region.

This article is organized as follows: Section \ref{sec:model} presents the Dirac-Bogoliubov-de Gennes Hamiltonian for TI surface, and the application of the asymptotic-solutions Green's functions method, for delimited surface regions with induced ferromagnetism or s-wave superconductivity. We also schematize how to couple these regions by using the Dyson equation to obtain the Green's functions and the transport observables of a system with multiple regions. In section \ref{sec:states}, it is discussed the concordance between our results for the NF and FS junctions and the known in the literature, principally on the induced topological phases and the associated interface bound states. In section \ref{sec:junctions}, the transport properties of systems with three regions are explored and discussed. Finally, in section \ref{sec:conc} our conclusions and perspectives are presented. 

\section{Model and transport observables \label{sec:model}}

This section is focused on the analysis of the energy spectrum and transport observables of systems that can be modeled as 2D junctions with different parallel interfaces and transversal translational invariance. The surface states of a strong TI of the Bi$_{2}$Se$_{3}$ family are described by an effective zero-mass Dirac Hamiltonian  at the $\boldsymbol{\Gamma}$ point \cite{Zhang_2009,Zhang_2010}. For the $+\mathbf{\hat{z}}$ surface it is given by $\mathcal{H}_s\left( \mathbf{r}\right) = 
v_F\left(\boldsymbol\sigma \times\mathbf{\hat{p}}\right) _{z}$, where $v_F$ is the speed of the charge carriers at the Fermi level,  $\mathbf{\hat{p}}=-i\mathbf{\hbar\boldsymbol{\nabla_\mathbf{r}}}$ is the momentum operator and $\boldsymbol\sigma=(\sigma_x,\sigma_y,\sigma_z)$ is the Pauli vector in the spin subspace.
First, was considered the case of a superconducting region (S) on the surface of a TI. There, the direct contact of the TI with a conventional superconductor favors the tunneling of Cooper pairs, inducing a superconducting state by the
surface proximity effect. For this region, the elementary excitations of the system are
described by the BdG-Dirac Hamiltonian with a spin-singlet $s$-wave order parameter
\begin{gather}
\mathcal{H}\left( \mathbf{r}\right)=\left( 
\begin{array}{cc}
\mathcal{H}_s\left( \mathbf{r}\right) -E_{F} & \Delta _{0}i\sigma
_{y} \\ 
-\Delta _{0}i\sigma _{y} & E_{F}-\mathcal{H}_s^{T}\left( -\mathbf{r}%
\right)%
\end{array}%
\right) \text{,} \label{HBdG}
\end{gather}
where $E_{F}$ is the surface Fermi energy. This Hamiltonian can also describe normal regions (N) by doing $\Delta_0=0$. If the surface of an N region contacts a ferromagnetic material with perpendicular magnetization vector $\mathbf{M}=M\mathbf{\hat{z}}$, a ferromagnetic region (F) is obtained. Hence, the surface Hamiltonian $\mathcal{H}_s\left(\mathbf{r}\right)$ in Eq.(\ref{HBdG}) acquires an additional Zeeman-type term of the form $\mathcal{H}_{Z}=\mathbf{M}\cdot\boldsymbol\sigma=M\sigma_z$ (in this work we consider $E_F=0$ for ferromagnetic regions to avoid possible transport channels through the ferromagnetic insulator bulk \cite{Nagaosa_2010}). 

By assuming translational invariance of the regions in $y$ direction, the eigenspinors of the Hamiltonian (\ref{HBdG}) have the form $\psi^{\mu }\left( x\right) \mathrm{e}^{iqy}$ where the superscript $\mu=qe,qh$ indicates electron- or hole-like quasiparticle solutions, and $q$ the conserved wave vector in $y$ direction. Then, the advanced and retarded Green's functions can be written as $\hat{%
	g}^{r,a}\left( E,x,x^{\prime },y-y^{\prime }\right) =\int dq\mathrm{e}%
^{iq\left( y-y^{\prime }\right) }\hat{g}^{r,a}\left( E,q,x,x^{\prime
}\right) /2\pi$, where Fourier transform satisfies the inhomogeneous equation 
\begin{equation}
\left[ \left(E\pm 0^{+}\right) -\mathcal{H}\left(x,q\right) %
\right] \hat{g}^{r/a}\left( E,q,x,x^{\prime }\right) =\delta
\left(x-x^{\prime }\right)\text{,} \label{GreenTI}
\end{equation}
with $E$ the excitation energy of the system and $0^{+}$ represents an infinitesimal scalar. 

The \textit{asymptotic solutions method} was implemented to obtain solutions of equation (\ref{GreenTI}) that satisfy specific boundary conditions for each region. This method has been used to find the Green's functions for the Sturm - Liouville equation \cite{Butkov}, the Schrödinger equation \cite{Linderberg}, the Bogoliubov-de Gennes equation in NS junctions (McMillan's formalism) \cite{McMillan_1968}, and recently in some graphene-based superconducting systems with finite-sized regions \cite{Herrera_2010,Gomez_2011,Casas_2019_1,Casas_2019}. In this method, the equilibrium Green's functions for each region are calculated from the scattering solutions of (\ref{HBdG}) at the boundaries as follows 
\begin{equation}
\hat{g}(x,x^{\prime })=\left\{ 
\begin{array}{cc}
\sum\limits_{\mu ,\nu =e,h}\hat{C}_{\mu \nu }\Psi _{L}^{\mu }\left( x\right) 
\tilde{\Psi}_{R}^{\nu T}\left( x^{\prime }\right) & x<x^{\prime } \\ 
\sum\limits_{\mu ,\nu =e,h}\hat{C}_{\mu \nu }^{\prime }\Psi _{R}^{\nu
}\left( x\right) \tilde{\Psi}_{L}^{\mu T}\left( x^{\prime }\right) & 
x>x^{\prime }%
\end{array}%
\right. \text{,}  \label{GreenZZ2}
\end{equation}
where $\Psi_{L,R}^{\mu }\left( x\right)$ are the asymptotic solutions of the region that obeys the boundary conditions at the left ($L$) or right ($R$) edge,
$\tilde{\Psi}_{L,R}^{\mu }\left( x\right)$ are the asymptotic solutions associated with $\mathcal{H}^{T}(x,-q)$ ($\tilde{\Psi}_{L,R}^{\mu }\left( x\right) =\mathcal{P}\Psi
_{L,R}^{\mu }\left( x\right)$ with $\mathcal{P}=I$ the inversion operator for this case), and $\hat{C}_{\mu \nu }$ are coefficient matrices determined by the equation (\ref{GreenTI}), as shown in Appendix \ref{sec:app1}, where we present the specific analytical expressions of the Green's functions implemented in the following sections.

\begin{figure}[h]
	\centering
		\includegraphics[width=1\columnwidth]{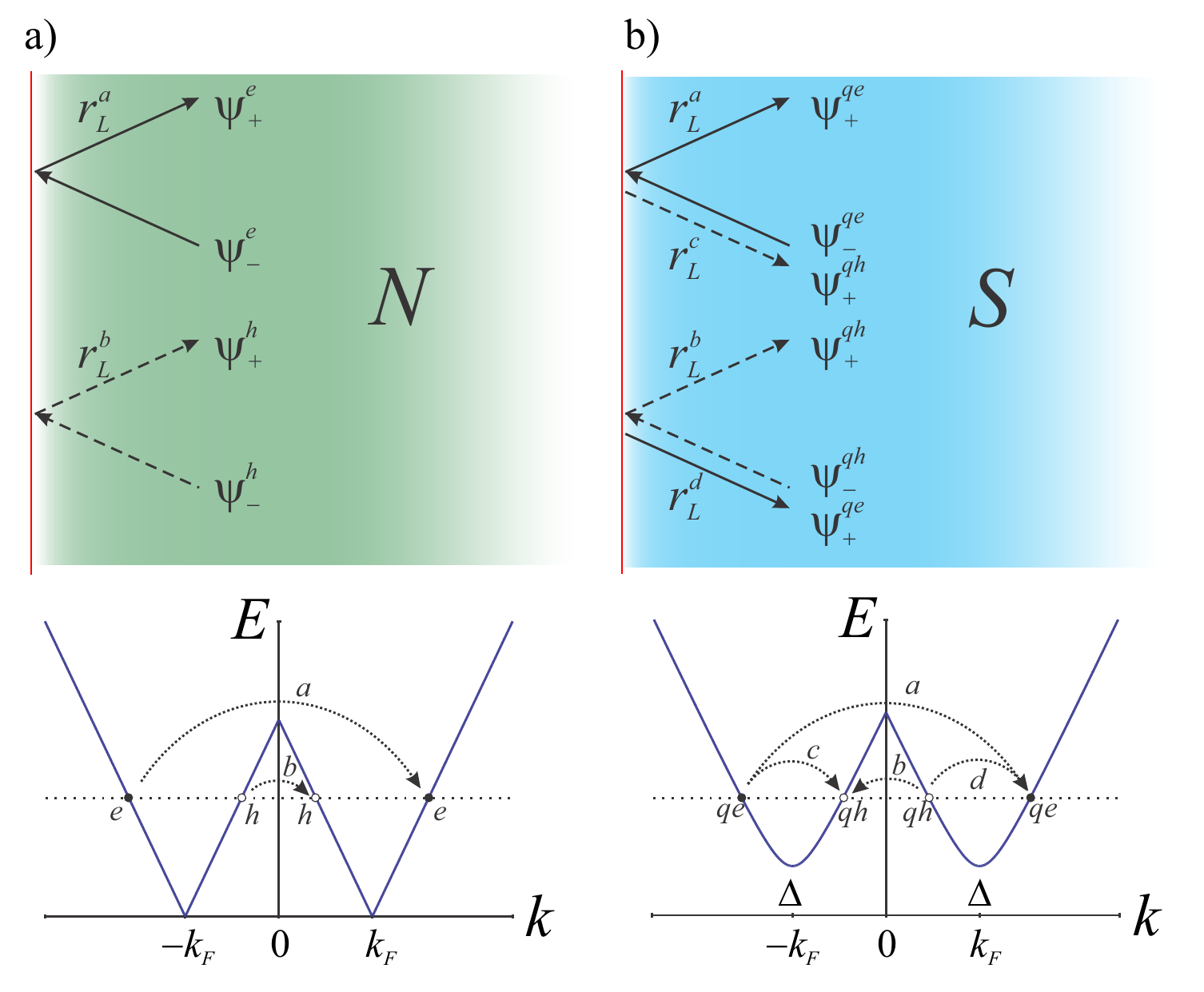}
		\caption{Scattering processes present at the left boundary of N and S regions. Solid (dashed) arrows represent group velocity directions of electrons (holes) in panel a), and group velocity directions of electron-like (hole-like) quasiparticles in panel b). The lower panels show the spectrum of each region and the different dispersion processes in a left edge. Curved arrows $a$ and $b$ indicate conventional reflections processes while arrows $c$ and $d$  illustrate branch crossing processes.
	}
	\label{fig:processes}
\end{figure}

For normal and ferromagnetic surfaces we have normal excitations [$\mu=e,h$ in equation (\ref{GreenZZ2})] and asymptotic solutions consist of conventional reflection processes at the boundaries, where the reflected particle is the same type as the incident particle [processes $a$ and $b$ in Fig. \ref{fig:processes}a)]
	\begin{eqnarray}
	\Psi _{L/R}^{e}\left( x\right) &=&\psi _{\mp}^{e}\left( x\right) +r_{L/R}^{a}\psi
	_{\pm}^{e}\left( x\right)\text{,} \label{asymp2}\\
	\Psi _{L/R}^{h}\left( x\right) &=&\psi _{\pm}^{h}\left( x\right) +r_{L/R}^{b}\psi _{\mp}^{h}\left( x\right)\text{,}  \notag 
	\end{eqnarray}
where $\psi _{\varepsilon }^{\mu }\left( x\right)$ are the eigenspinors of (\ref{HBdG}) with $\Delta_0=0$ propagating in $\varepsilon \mathbf{\hat{x}}$ direction, and the $r_{L/R}^{i}$ with $i=a,b$ are the reflection coefficients at left ($L$) or right ($R$) edge defined by the boundary conditions (see Appendix \ref{sec:app1A} for details).	

For a superconducting region, asymptotic solutions include besides quasiparticle reflection processes, branch crossing processes (due to the fast variation of the pair potential near the boundary), where the incident and reflected quasiparticles are of different type [processes $c$ and $d$ in Fig. \ref{fig:processes}b)]. Then these can be written as
\begin{eqnarray}
\Psi _{L/R}^{qe}\left( x\right) &=&\psi _{\mp}^{qe}\left( x\right) +r_{L/R}^{a}\psi
_{\pm}^{qe}\left( x\right) +r_{L/R}^{c}\psi _{\mp}^{qh}\left( x\right) \text{,} \label{asymp}\\
\Psi _{L/R}^{qh}\left( x\right) &=&\psi _{\pm}^{qh}\left( x\right) +r_{L/R}^{b}\psi _{\mp}^{qh}\left( x\right) +r_{L/R}^{d}\psi _{\pm}^{qe}\left( x\right) 
\text{,}  \notag 
\end{eqnarray}%
where $\psi _{\varepsilon }^{\mu }\left( x\right)$ (with $\mu=qe,qh$) are the eigenspinors of (\ref{HBdG}) and $r_{L/R}^{i}$ with $i=a,b,c,d$ are the quasiparticle reflection coefficients at $L/R$ edge (see Appendix \ref{sec:app1B} for details). On the other hand, for semi-infinite regions, the open boundary condition at a given side implies no reflected contributions, as summarized in Table \ref{tab:nullcoeff}.
\begin{table}[h]
	\centering
	\begin{tabular}{c|c|c}
		\hline\hline
	    \diaghead{\theadfont Extensionnnnnnnn}{Extension}{Type}& Normal/magnetic & superconducting \\ \hline
		\thead{semi-infinite \\ (left)} & 
		\thead{$r_{L/R}^{c}$, $r_{L/R}^{d}$, \\ $r_{L}^{a}$, $r_{L}^{b}$}
		& \thead{$r_{L}^{c}$, $r_{L}^{d}$, \\ $r_{L}^{a}$, $r_{L}^{b}$} \\ \hline
		finite & 
		\thead{$r_{L/R}^{c}$, $r_{L/R}^{d}$}
		& none \\ \hline
		\thead{semi-infinite  \\  (right)} & 
		\thead{$r_{L/R}^{c}$, $r_{L/R}^{d}$, \\ $r_{R}^{a}$, $r_{R}^{b}$}
		& \thead{$r_{R}^{c}$, $r_{R}^{d}$, \\ $r_{R}^{a}$, $r_{R}^{b}$} \\ \hline\hline
	\end{tabular}%
	\caption[]{Null reflection coefficients for the distinct types of regions studied.}
	\label{tab:nullcoeff}
\end{table}

Since TI's surface lacks of borders, it is necessary to introduce artificial boundary conditions for each region provided that perfect transparency is recovered when coupling different regions. To simplify the calculations, we adopted artificial boundary conditions for the spin simulating opposite infinite magnetic barriers in $x=x_L$ and $x=x_R$ (analogous to those of a graphene ribbon with zigzag edges along the $y$ axis \cite{Herrera_2010}). Here, we adopted the following choice for the boundary conditions

\begin{equation}
\Psi _{L}^{\mu }\left( x_{L}\right) |_{\downarrow }=\Psi
_{R}^{\mu }\left( x_{R}\right) |_{\uparrow }=0.\label{zzcond}
\end{equation}

\begin{figure}[h]
	\centering
	\includegraphics[width=1\columnwidth]{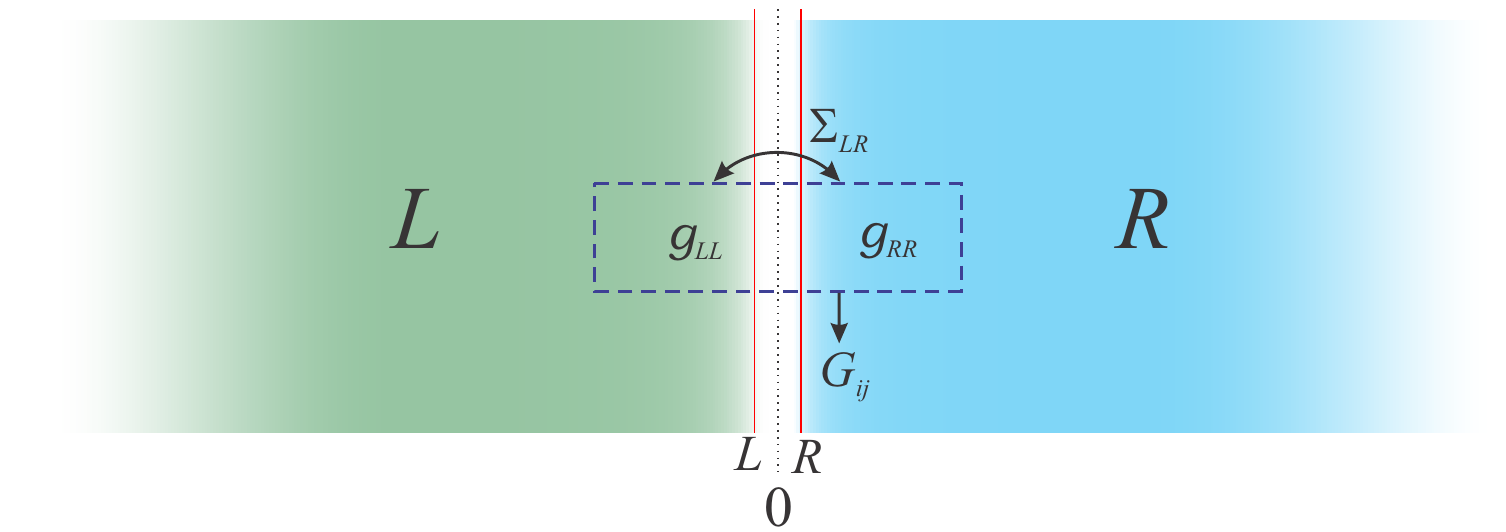}
	\caption{Schematic representation of the calculation of the equilibrium Green's functions of a junction with two semi-infinite regions. The Green's functions of the $L$ and $R$ regions (calculated with the asymptotic solutions method) are coupled with the Dyson equation (Details in Appendix \ref{sec:app1C}). } \label{fig:regiones}
\end{figure} 
According to the selected boundary conditions, the coupling of different regions placed in series is modeled by the Hamiltonian approach \cite{Yeyati_1996}. There, the microscopic hopping of charge carriers between available channels at the edges of adjacent regions (Fig. \ref{fig:regiones}) is described by a tight-binding Hamiltonian of the form \cite{Yeyati_1996,Herrera_2010,Gomez_2011,Casas_2019_1,Casas_2019}
\begin{align}
H_{T}& =t\int \mathrm{d}q
\hat{c}_{q,L\downarrow}^{\dagger }\hat{c}_{q,R\uparrow} +\text{h.c.,}
\label{eq:hopping}
\end{align}%
where $t=\hbar v_{F}$ is the hopping amplitude associated with the perfect (or transparent) coupling between regions on the TI’s surface, and $\hat{c}_{q,i\sigma }$ are the annihilation operators for charge carriers at the edge of the $i =L,R$ region with wave number $q$ and spin projection $%
\sigma = \uparrow ,\downarrow $. Consequently, given the equilibrium Green's functions of the two adjacent regions $\hat{g}_i=\hat{g}(x_i,x^{\prime}_i)$, as those defined above, it is possible to calculate the equilibrium Green's function $\hat{G}_{ij}=\hat{G}(x_i,x^{\prime}_j)$ of the entire system by
using a Dyson equation of the form \cite{Yeyati_1996,Herrera_2010,Gomez_2011,Casas_2019_1,Casas_2019} 
\begin{equation}
\hat{G}_{ij}=\hat{g}_{ij}+\hat{g}_{ik}\hat{\Sigma}_{kl}\hat{G}_{lj}%
\text{,} 
\label{eq:Dyson_equation}
\end{equation}
where $\hat{g}_{ij}=\hat{g}_{i}\delta_{ij}$ and $\hat{\Sigma}_{ij}$ ($i\neq j$) are the coupling `self-energies' between the adjacent edges of the two regions and are given by the matrix form of the hopping Hamiltonian (\ref{eq:hopping}) 
\begin{eqnarray}
\hat{\Sigma}_{LR}
&=&\hat{\Sigma}_{RL}^{T}=t\tau _{z}(\sigma _{x}-i\sigma _{y})/2\text{.}  \label{auto3}
\end{eqnarray}

The formal details in the implementation of the Dyson equation for junctions with two or more coupled regions are illustrated in Appendix \ref{sec:app1C}.

Once the equilibrium Green's functions of the system have been calculated with the Dyson equations, the momentum-resolved spectral density $A(x,E,q)$ and the density of states
(DOS) $\rho \left( x,E\right) $ are given by the standard relations
\begin{eqnarray}
A(x,E,q) &=&-\frac{1}{\pi }\text{\textrm{Im}}\left\{ \mathrm{Tr}\hat{G}%
_{ee}^{r}\left( x,x,E,q\right) \right\} \text{,} \\
\rho \left( x,E\right) &=&\int \mathrm{d}qA(x,E,q)\text{.}
\end{eqnarray}

In this work, we analyze the transport properties of some junctions with two and three different regions. The differential conductance of a system such as the presented in Fig. \ref{fig:regiones} is given by $\sigma=\partial I/\partial V$, with $V$ the applied bias voltage and $I$ the stationary current through the junction \cite{Yeyati_1996,Gomez_2011,Casas_2019_1,Casas_2019} 
\begin{equation}
I=\frac{e}{2h}\int \mathrm{d}q\mathrm{d}E\mathrm{Tr}\left( \tau _{z}\left[ 
\hat{t}\hat{G}_{q,RL}^{+-}(E)-\hat{t}^{\dag }\hat{G}_{q,LR}^{+-}(E)%
\right] \right) \text{,} \label{current}
\end{equation}
where $\hat{t}\equiv\hat{\Sigma}_{LR}$ and the $\hat{G}_{q,ik}^{+-}(E)$ are the non-local Keldysh (or non-equilibrium) Green's functions evaluated at the edges of the $L$ and $R$ regions, which are related to the equilibrium Green's functions of the system as shown in detail in Appendix \ref{sec:app2} for a three-region system.

\section{Examples: two-region junctions\label{sec:states}} 
To illustrate the implementation and validity of our method, two systems previously studied in the literature were considered, the FF and FS junctions. The emphasis is focused on the interface states and momentum-resolved spectral density. The details of the calculations are presented in Appendix \ref{sec:app1C}. 
\subsection{FF junction}
\begin{figure}
	\centering
	\includegraphics[width=1\columnwidth]{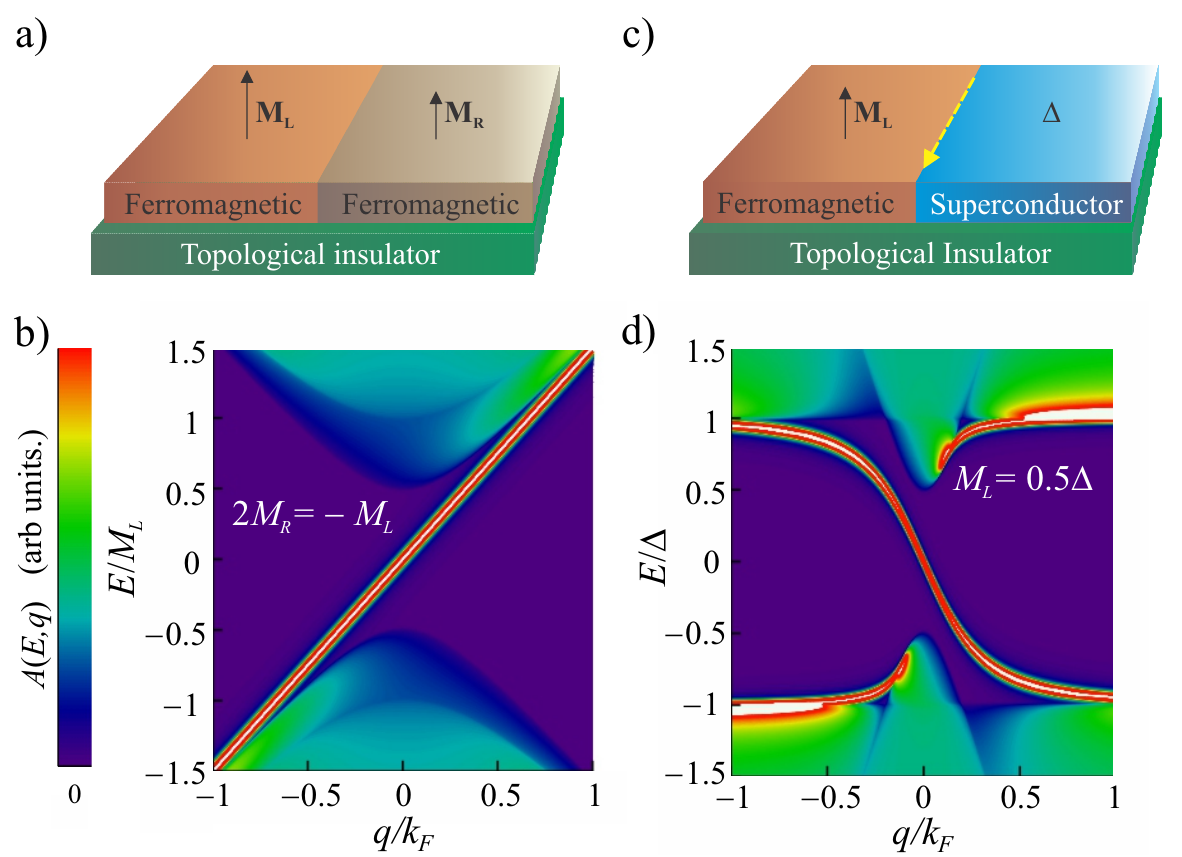}
	\caption{a) A Bi$_{2}$Se$_{3}$ block in interfacial contact with two adjacent ferromagnetic materials. b) Momentum-resolved spectral density at the interface of the junction depicted in panel a) for antiparallel magnetizations ($E_{FL}=E_{FR}=0$). c) A Bi$_{2}$Se$_{3}$ block in interfacial contact with a ferromagnetic insulator and a conventional superconductor. The yellow dashed arrow indicates the direction of the chiral Majorana IBS. d) Momentum-resolved spectral density at the interface of the junction depicted in panel c) for $M_L=0.5\Delta$ ($E_{FL}=0$ and $E_{FL}=100\Delta$).} \label{fig:TI_MM}
\end{figure}

First, the case of an FF junction was considered, where the surface of a topological insulator is placed in contact with two adjacent ferromagnetic insulators with polarized magnetization in $z$ direction as illustrated in Fig. \ref{fig:TI_MM} a).
In this system, a magnetization perpendicular to the TI surface induces a gap in the energy spectrum of the system and now the surface state constitutes the QAHE topological phase. The surface exhibits a chiral edge state at a magnetic domain wall with associated Hall conductance $\sigma _{y}= \mathrm{sgn}\left( M_{R}\right)
e^{2}/h$. In turn, this conductance is proportional to the Chern number, a topological invariant for symmetry class A \cite{Zhang_2008, Nagaosa_2010_F, Chang_2013, Xu_2014, Brey_2014}.

This system can be modeled as a junction between two
semi-infinite ferromagnetic regions perfectly coupled at $x_{0}=0$. The
Green's function of the compound system at the interface is obtained
by the Dyson equation (\ref{eq:Dyson_equation}) [see Eq.(\ref{DY1})]. The Green's functions poles contain
information of the bound states present in the FF interface. There are no subgap solutions for zero doping in both regions and $M_{R}=M_{L}$, while for $M_{R}=-M_{L}$ (domain wall configuration) we obtain the linear dispersion relation $
E=\mathrm{sgn}\left( M_{R}\right) \hbar v_{F}q 
$ for the QAHE chiral edge states. The interface momentum-resolved spectral density for this configuration is shown in panel b) of Fig. \ref{fig:TI_MM}. This illustrates that our approach leads to well-known results in the literature \cite{Zhang_2008,Nagaosa_2010_F,Brey_2014}. At the same time, it allows the derivation of the dispersion relation of the interface bound states, as well as the direct calculation of the momentum-resolved spectral density at the interface.
 
\subsection{FS junction}
Now the case of a FS junction is considered. There, the right ferromagnetic insulator of the FF junction is replaced by a conventional $s$-wave superconductor as shown in Fig. \ref{fig:TI_MM} c). At the weak-coupling limit, the proximity effect between an s-wave superconductor and the TI surface gives rise to an effective spinless $p_x+ip_y$ superconducting order parameter. Here, the Andreev bound states at FS interfaces are chiral Majorana modes \cite{Kane_2008, Xu_2015,Sun_2016_TI, Sun_2017}. The local Green's function of the coupled system at the interface is also given by (\ref{DY1}) with the right unperturbed Green's function presented in (\ref{superR}).  
The poles of this coupled Green's function lead to the dispersion relation of a chiral Majorana mode [Fig. \ref{fig:TI_MM} d)] 
\begin{equation}
E\left( M_{L},q\right) =\frac{-\mathrm{sgn}\left( M_{L}\right) \left\vert
	\Delta \right\vert \hbar v_{F}q}{\sqrt{\left( \left\vert \Delta \right\vert
		+M_{L}\right) ^{2}+\left( \hbar v_{F}q\right) ^{2}}}\text{.}
\label{ChiralMajorana}
\end{equation}

The information associated with the chirality of this Majorana IBS is contained in factors of the full spectral density, as in the case of IBS in graphene \cite{Casas_2019_1}. In panel d) of Fig. \ref{fig:TI_MM} there are a couple of subgap interface states with opposite chirality. These states correspond to the remaining IBS modes with energy $%
E\left( -M_{L},q\right) $, which are suppressed in the magnetic gap region. This occurs due to the chiral effect of the magnetization direction in the spin polarization of the helical surface states. Again, the formalism implemented here leads to results reported in the literature \cite{Kane_2008,Tanaka_2009,Nagaosa_2010}.

\section{THREE-REGION JUNCTIONS: GEOMETRICAL EFFECTS \label{sec:junctions}}
In this section we analyze the local spectral density and differential conductance of junctions with a finite intermediate region, by following the Green's function approach exposed in the previous sections. In these systems, the finite size of the central region results in the appearance of Fabry-Pérot resonances that are manifested in the transport properties of the system. 
\subsection{NFN junction}
\begin{figure}[h]
	\centering
	\includegraphics[width=1.0\columnwidth]{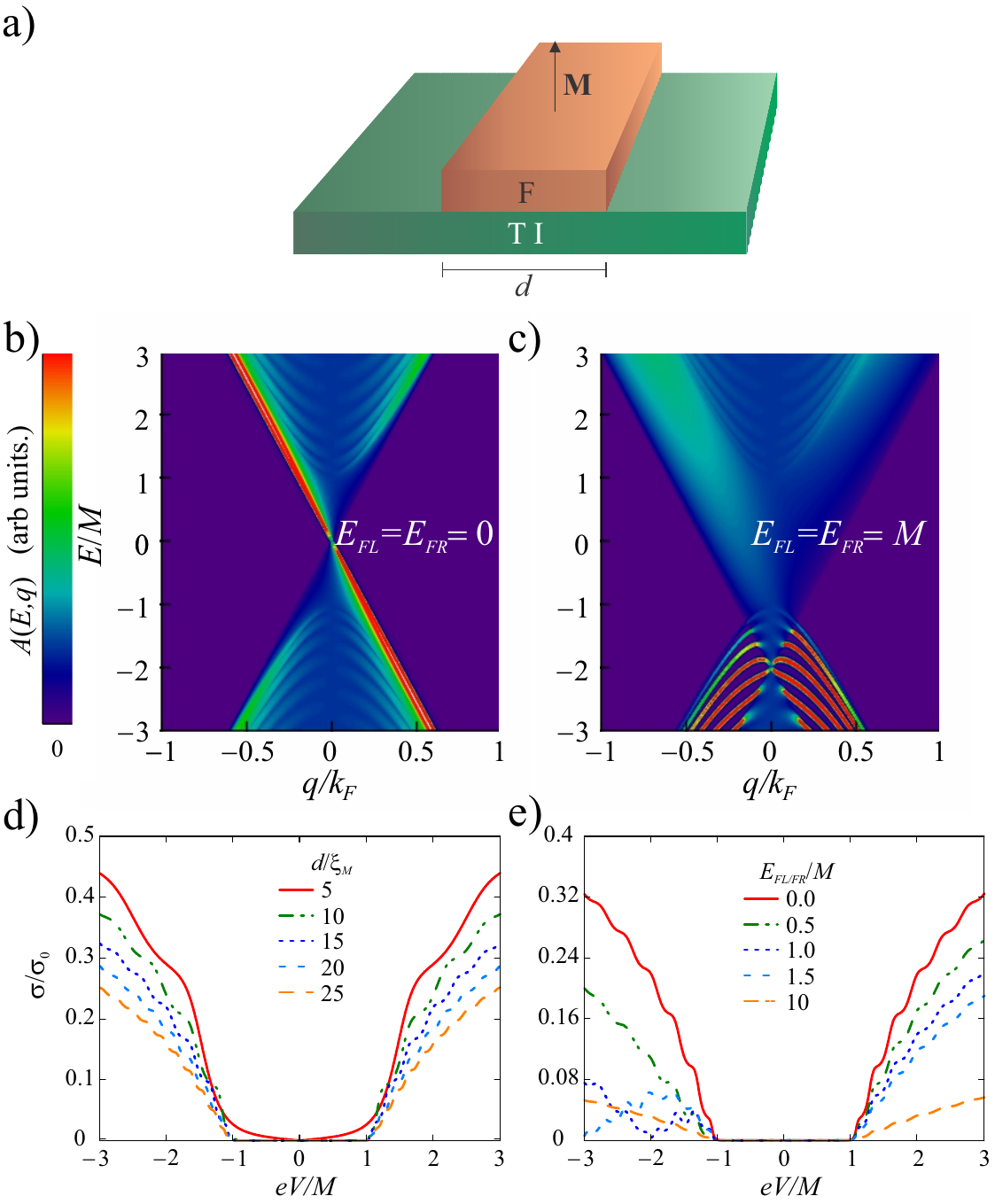}
	\caption{a) A Bi$_{2}$Se$_{3}$ block in interfacial contact with a ferromagnetic insulator of finite width $d$. Panels b) and c) show the momentum-resolved spectral density of the junction evaluated at the FN interface for two different doping levels of the lateral regions. NFN differential conductance for d) different widths $ d $ of the ferromagnetic region, e) different doping levels of the lateral regions ($\xi_{M}\equiv\hbar v_F/M$).}\label{fig:TI_NMN}
\end{figure}

The NFN junction consists of an infinite surface of a topological insulator with a ferromagnetic region of finite width $d$ as illustrated in Fig. \ref{fig:TI_NMN} a). 
The spectral density of the system at the FN interface ($x_{0}=0$) is similar to the case of an infinite NF junction for small doping of the normal lateral regions ($E_{F,L/R}/M\sim 0$). It exhibits a Dirac cone with vertex at $E=-E_{FC}=0$ and a chiral edge state associated with the QAHE [Fig. \ref{fig:TI_MM} b)]. However, the spectral density presents a pattern of faint \textquotedblleft parabolic\textquotedblright undulations inside the cone from $\left\vert E\right\vert >\left\vert
M\right\vert $, whose number increases in proportion to the width of the ferromagnetic region, evidencing its geometric character. These
undulations correspond to weak quasi-bound modes or Fabry-P\'erot resonances (FPRs) that occur within the ferromagnetic region due to specular reflection processes at the interfaces. The introduction of the mass term in the Dirac Hamiltonian (associated with the magnetization vector perpendicular to the surface) reduces the transmission probability and introduces reflected modes at the interfaces \cite{Setare_2010,Xie_2017}, even though Klein tunneling ensures perfect transmission at normal incidence between normal surfaces \cite{Xie_2017,Lee_2019}. Since these modes propagate in $x$ direction, they appreciably contribute to the differential conductance of the junction as shown in Fig. \ref{fig:TI_NMN} d) for several widths of the magnetic region. Besides, this is suppressed in the range $\left\vert eV\right\vert <\left\vert M\right\vert$ due to the absence of transport channels inside the magnetic gap, and the suppression effect is proportional to the size of the ferromagnetic region. Also, the conductance exhibits a series of undulations in the regions $\left\vert eV\right\vert >\left\vert
M\right\vert $ due to the formation of quasi-bound modes.

Even though the chiral edge state characteristic of zero doping cases disappears for non-zero doping of the lateral regions [Fig. \ref{fig:TI_NMN} c)], the spectral density still retains a chiral character. Besides, it has two overlapping cones: the first is a Dirac cone with a vertex at $E=-E_{F,L/R}$ that correspond to the normal lateral regions, and the second is a `gapped cone' associated with the ferromagnetic central region. The later presents a pattern of parabolic undulations for $%
E>0 $ (as in the case with zero doping), while for $E<0$ presents a series of clearly
defined parabolic FPR bands that fade when entering the Dirac
cone of the lateral regions. The FPRs significantly contribute to
transverse transport as shown in Fig. \ref {fig:TI_NMN} e). The figure illustrates the differential conductance for various doping levels of the ferromagnetic region. It is observed that the electrodes conical DOS exhibits a zero conductance minimum around $eV=-E_{F,L/R}$, and the increase of the lateral doping levels induces a global reduction in conductance due to the chosen normalization (See Appendix \ref{sec:app2} for details).

\subsection{NFS junction}
\begin{figure}[h]
	\centering
	\includegraphics[width=1.0\columnwidth]{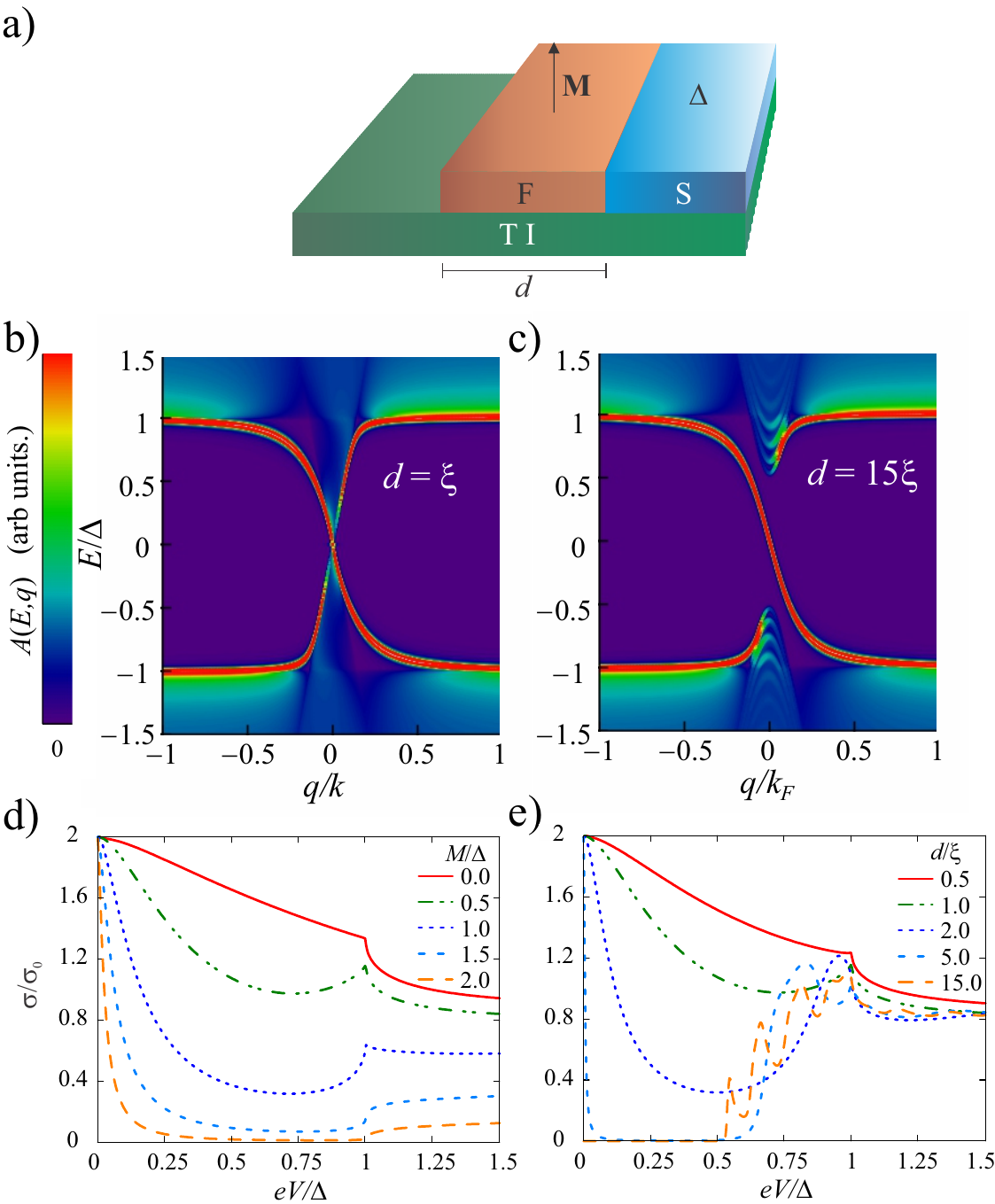}
	\caption{a) A Bi$_{2}$Se$_{3}$ block in interfacial contact with a ferromagnetic insulator of finite width $d$ and a conventional superconductor ($E_{FL}=0$ and $E_{FR}=100\Delta$). Panels b) and c) show the momentum-resolved spectral density of the junction at the FS interface for two different widths $d$ and $E_{FC}=0$. Differential conductance of the NFS junction for d) different values of the magnetization and $d\sim \xi$; e) for different widths of the central region and $M=0.5\Delta$ ($\xi\equiv \Delta/\hbar v_F$).}
	\label{fig:TI_NMS}
\end{figure}
In the case of the NFS junction, the normal region at the right of the NFN junction is placed in contact with an
$s$-wave superconductor as seen in Fig. \ref{fig:TI_NMS}, panel a). The spectral density at the FS interface is
presented in panels b) and c). For a width $d\sim \xi $ [Fig.\ref{fig:TI_NMS} b)], the system exhibits a pair of subgap bands, a negative-slope band corresponding to the chiral Majorana mode (\ref{ChiralMajorana}) and a positive-slope band associated with the IBS solution $E\left( -M,q\right) $. The later
is attenuated in the vicinity of $q\sim 0$ due to the selective effect of the magnetization vector in the spin polarization of the helical surface states. This situation is analogous to that found for the FS junction of the previous section, except for the attenuation region that turns smaller due to the finite size of the ferromagnetic region. In
contrast, for $d=15\xi$ [Fig.\ref{fig:TI_NMS} c)] there is an evident attenuation region in the range $\left\vert E\right\vert <\left\vert M\right\vert $ as in the case of the FS junction. In this case, there is a pattern of undulations inside the paraboloid (with gap $2M$) associated with the central region.

Regarding the longitudinal conductance for low doping levels of the left electrode, a ferromagnetic region with $d\sim \xi$ results in a reduction of transport proportional to the induced magnetization in the range $0<\left\vert eV\right\vert<\left\vert \Delta \right\vert $ [Fig. \ref{fig:TI_NMS} d)]. A zero-energy conductance peak (ZBCP) is preserved due to the presence of a chiral Majorana mode at the NS interface. However, this state rapidly decays in the $x$ direction, and for $d>5\xi$, the ZBCP begins to drop whereas some undulations start to emerge for $\left\vert
eV\right\vert >\left\vert M\right\vert $. These oscillations are associated with the weak FPR bands inside the paraboloid due to the finite size of the ferromagnetic region [\ref{fig:TI_NMS} e)]. 

\subsection{FNS junction}
\begin{figure}[]
	\centering
	\includegraphics[width=1\columnwidth]{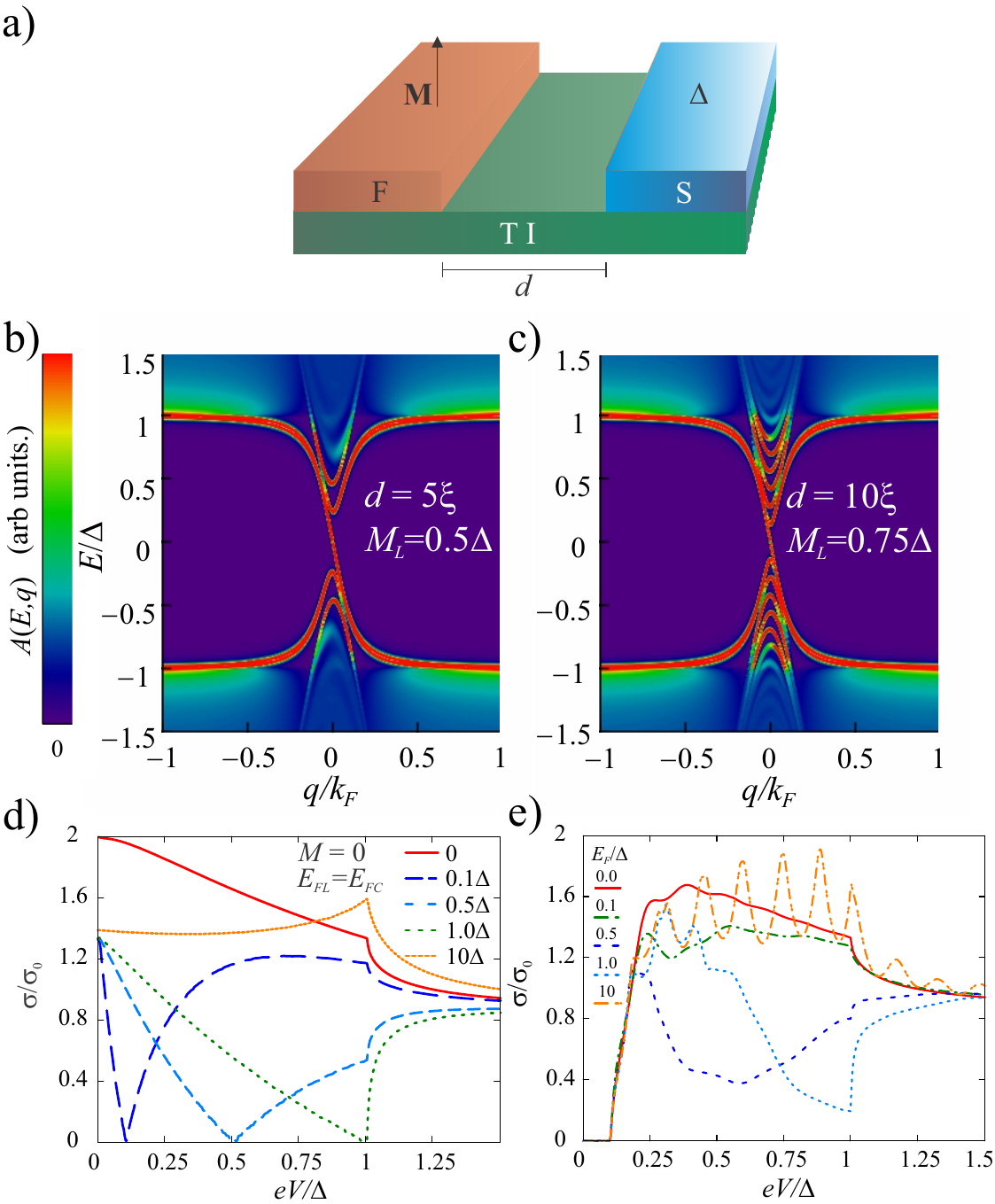}
	\caption{a) A Bi$_{2}$Se$_{3}$ block in interfacial contact with a ferromagnetic insulator and a conventional superconductor separated at a distance $d$ ($E_{FL}=0$ and $E_{FR}=100\Delta$). Panels b) and c) present the momentum-resolved spectral density of the junction evaluated at the FS interface for two different values of width and magnetization and $E_{FC}=0$. d)  Differential conductance of the NS junction for various doping levels of the N region ($E_{FR}=100\Delta$), e) differential conductance of the FNS junction for various doping levels of the N region ($M_{L}=0.1\Delta$ and $E_{FR}=100\Delta$).}
	\label{fig:TI_MNS}
\end{figure}

Finally, we consider the case of the FNS junction shown in Fig. \ref{fig:TI_MNS} a). Panels b) and c) of this figure show the spectral density evaluated at the NS interface for two different values of $d$ and $M_{L}$ ($E_{FC}=0$), where the chiral IBSs at the FS interface are observed, including the chiral Majorana mode of $E\left(M_{L},q\right) $ (\ref{ChiralMajorana}), now accompanied by FPR bands originated by the formation of Andreev quasi-bound states inside the central normal region. These bound states are the result of the constructive superposition of propagating states scattered at the interfaces of the central region. Hence, incoming electron-like quasiparticles from the left ferromagnetic electrode are Andreev-reflected as hole-like quasiparticles at the NS interface, and are partially transmitted and reflected at the FN interface, depending on the angle of incidence and the value of $M$ (equivalently for incoming hole-like states reflected as electron-like states at the NS interface). 

As noted in the previous case, these FPR bands are attenuated outside the gap and  its number is proportional to the width of the
central region (as in the case of the Andreev quasi-bound states present in NINS junctions \cite{Lofwander_2001,Casas_2019_1}). However, in this case, the transmission at normal incidence (associated with Klein tunneling) is not reduced by an insulating contact or an imperfect coupling, but due to the presence of the magnetic-mass term in the surface Hamiltonian. This effect gives rise to reflected waves at the FN interface, even at normal incidence as stated above for the NFN junction. All this geometrical effects could be observed in spectroscopy experiments as ARPES or STS.

In the absence of magnetization and for $E_{FC}=E_{FL}$, the system becomes an NS junction and its differential conductance presents the specular Andreev reflections profile. This is characterized by a zero conductance minimum located at $eV=-E_F$ that corresponds with the minimum of the DOS associated with the Dirac point of the N region \cite{Beenakker_2006} [Fig.\ref{fig:TI_MNS} d)]. Nevertheless, for the FNS junction case, the presence of the ferromagnetic insulator at the surface of the left electrode, with a finite normal region in the middle, has interesting effects in the longitudinal conductance. 

First, the longitudinal differential conductance [Fig.\ref{fig:TI_MNS} e)] shows zero conductance region $|eV|<|M|$ for all the doping levels due to the absence of transport channels inside the magnetic gap of the left semi-infinite electrode. As mentioned in the section above, the suppressor effect of $M$ is proportional to the size of the ferromagnetic region and is total for an infinite-size electrode. Second, there is a minimum at $eV=-E_{F}$ associated with the vertex of the Dirac cone of the central region, that separates specular and retro-Andreev reflection regimes [Fig. \ref{fig:TI_MNS} d)]. However, in this case this minimum is attenuated due to the contribution of some FPR bands at the Fermi level of the N region. Third,  for high doping levels, parabolic FPR bands manifest in the conductance as a series of small peaks whose number is proportional to the width of the central N region. As in a graphene-based NINS junction \cite{ZYZhang_2008,Casas_2019_1}, these conductance resonances have a higher intensity when increasing the doping of the central region as shows Fig. \ref{fig:TI_MNS} e). This effect is because by increasing the doping of the central normal region, the reflection processes at the NS interface transit from the specular to the retro-Andreev reflections regime, that, in a semi-classical perspective favors the formation of closed paths for Andreev bound  states \cite{Lofwander_2001,ZYZhang_2008}.  

\section{Conclusion \label{sec:conc}}

In this paper, we adapted the asymptotic-solutions Green's functions approach for the case of junctions on the surface of strong TIs with ferromagnetic and $s$-wave superconducting interfaces. This method has been successfully implemented for 2D systems with multiple coupled regions like graphene-based superconducting junctions. For the construction of the Green's functions of each region, artificial boundary conditions were adopted, so that all the junction interfaces had perfect transparency when coupled with the Dyson equation. This method allows the study of the transport properties of a wide variety of junctions with the same basic elements. Besides, it also permits the direct calculation of the momentum-resolved spectral densities at the interfaces, which could be of interest for the identification and analysis of topological interface bound states.

The results obtained for junctions of two regions are consistent with those found in literature. In our study, the FF junction presents the characteristic chiral bound state of the QAHE for opposite magnetizations, and a gapped spectrum for parallel magnetizations. On the other hand, for the FS junction the chiral IBS and Majorana modes were found. The properties of some junctions with three regions were also studied. Regarding the NFN junction, the QAHE chiral edge state is observed for $E_F=0$ along with some FPRs. These resonances are originated in the reflection process at the interfaces, and their number is proportional to the width of the central region. With respect to the NFS junction, IBS and Majorana chiral modes are observed in addition to some weak FPRs. Respect to the longitudinal transport of this junction, the increase in magnetization reduces the subgap conductance except by a small peak for $eV=0$ associated with the chiral Majorana mode, which finally decays for widths of the central regions greater than the superconducting coherence length, while the number of peaks associated with the FPRs becomes relevant.

Respect to the FNS junction, this also presents chiral IBS and Majorana modes at the NS interface for zero doping of the central region, in addition to some strong subgap FPR bands associated with the Andreev bound states confined by conventional and Andreev reflection process at the interfaces. The number of these resonances also increases with the width of the central region and the associated conductance peaks become appreciable for high doping levels of the central region. On the other hand, due to the large size of the ferromagnetic left electrode, the conductance is suppressed for $|eV|<|M|$. For low doping of the central region, the subgap conductance structure resembles the characteristic profile of specular Andreev reflection of an NS junction in Dirac 2D systems. Finally, we expect this method to become a useful tool in further studies on electronic and transport properties of junctions and nanostructures on the TIs surface and other 2D similar systems, specially for the sake of the experimental identification and characterization of new topological phases and interface bound states present in these types of structures, either through ARPES or transport measurements.

\acknowledgments 
We acknowledge funding from COLCIENCIAS, Grant No. 110165843163 and Doctorate Scholarship 617.



\appendix

\section{Green's functions for uncoupled regions\label{sec:app1}}
In this appendix, we present the calculation of the equilibrium Green's functions of the uncoupled normal-ferromagnetic and superconducting regions by the asymptotic solutions method. Then, these Green functions can be coupled by using the Dyson equation to obtain the equilibrium Green's function of a system composed of several regions, as illustrated in Appendix \ref{sec:app1C}.
\subsection{Normal and ferromagnetic regions\label{sec:app1A}}
First, to calculate the asymptotic solutions of normal and ferromagnetic regions, it was necessary to find the eigenvalues and eigenvectors of Hamiltonian (\ref{HBdG}) for $\mathbf{M}=M\mathbf{\hat{z}}$ and $\Delta_0=0$. The spectrum is given by
\begin{equation}
E_{e/h}=\pm \left( \sqrt{\left( \hbar v_{F}\left\vert \mathbf{k}\right\vert
	\right) ^{2}+M^{2}}-E_{F}\right) \text{,}  \label{espectroM}
\end{equation}%
with eigenspinors
\begin{eqnarray}
\psi _{\varepsilon }^{e}\left( \mathbf{r}\right) =\mathrm{e}^{iqy}\mathrm{e}%
^{\varepsilon ik_{e}x}\left( \hat{\varphi}_{\varepsilon }^{e},0\right) ^{T}%
\text{,}\\
\psi _{\varepsilon }^{h}\left( \mathbf{r}\right) =\mathrm{e}%
^{iqy}\mathrm{e}^{\varepsilon ik_{h}x}\left( 0,\hat{\varphi}_{\varepsilon
}^{h}\right) ^{T}\text{,}
\end{eqnarray}%
where $\varphi _{\varepsilon }^{\mu}\left( \mathbf{r}\right)$ (with $\mu =e,h$) are the eigenspinors of the electrons/holes matrix sectors of Hamiltonian (\ref{HBdG}) 
\begin{eqnarray}
\hat{\varphi}_{\varepsilon }^{e} &=&\left( M_{+}^{e},-\varepsilon iM_{-}^{e}%
\mathrm{e}^{\varepsilon i\alpha _{e}}\right) ^{T}/\sqrt{2}\text{,}
\label{espinor1} \\
\hat{\varphi}_{\varepsilon }^{h} &=&\left( \varepsilon iM_{+}^{h}\mathrm{e}%
^{\varepsilon i\alpha _{h}},M_{-}^{h}\right) ^{T}/\sqrt{2}\text{,}
\label{espinor2} \\
M_{\pm }^{e} &=&\sqrt{E+E_{F}\pm M}/\sqrt{E+E_{F}}\text{,} \\
M_{\pm }^{h} &=&\sqrt{E_{F}-E\pm M}/\sqrt{E_{F}-E}\text{,} \\
\mathrm{e}^{i\alpha _{\mu }} &\equiv &\hbar v_{F}\left( k_{\mu }+iq\right)
/\left( E_{F}\pm E\,\right) \text{,}
\end{eqnarray}%
with $\mu =e,h$ and wave number in $x$
\begin{equation}
k_{e/h}=\mathrm{sgn}\left( E_{F}\pm E\right) \sqrt{\frac{\left( E_{F}\pm
		E\right) ^{2}-M^{2}}{\hbar ^{2}v_{F}^{2}}-q^{2}}\text{,}
\end{equation}%
where the sign-function sets the correct sign for the valence band. For the adopted `zigzag-type' artificial boundary conditions for spin $\psi _{L}^{\mu }\left( x_{L}\right) |_{\uparrow /\downarrow }=\psi
_{R}^{\mu }\left( x_{R}\right) |_{\downarrow /\uparrow }=0$, the reflection coefficients in (\ref{asymp}) are given by
\begin{eqnarray}
r_{R,\uparrow /L,\downarrow }^{a/c} &=&-\mathrm{e}^{2ik_{e/h}x_{R/L}}\text{,}\\
r_{L,\downarrow /R,\uparrow }^{a/c}&=&\mathrm{e}^{-2i\alpha _{e/h}}%
\mathrm{e}^{-2ik_{e/h}x_{L/R}}\text{,}\\
r_{L,\uparrow /R,\downarrow }^{a/c} &=&-\mathrm{e}^{2ik_{e/h}x_{L/R}}\text{,}\\
r_{R,\downarrow /L,\uparrow }^{a/c}&=&\mathrm{e}^{2i\alpha _{e/h}}\mathrm{%
	e}^{2ik_{e/h}x_{R/L}}\text{.}
\end{eqnarray}

Integrating the equation (\ref{GreenTI}) in $x$ over an infinitesimal region around $x^{\prime }$ we obtain the following auxiliary relation
\begin{equation}
\hat{g}\left( x^{\prime }+0^{+},x^{\prime }\right) -\hat{g}\left( x^{\prime
}-0^{+},x^{\prime }\right) =\frac{i}{\hbar v_{F}}\left( \tau _{z}\otimes
\sigma _{y}\right) \text{,}  \label{LigaduraTI}
\end{equation}
which allows to obtain the coefficient matrices $\hat{C}_{\mu\nu}$. In this case, by grouping similar terms in the constraint condition (\ref{LigaduraTI}) we found that the only non-zero coefficient matrices are $\hat{C}_{\mu \mu }=\hat{C}_{\mu \mu }^{\prime }$ (since there is not coupling between electrons and holes), and if we assume electron-hole symmetry  ($\hat{C}_{ee}=\hat{C}_{hh}$) we
have
\begin{equation}
\hat{C}_{ee}=\frac{-iN_{e}}{2\hbar v_{F}\text{\textrm{cos}}\alpha _{e}}%
\frac{\left( \tau _{0}+\tau _{z}\right) }{1-r_{R}^{a}r_{L}^{a}}+\frac{-iN_{h}}{2\hbar v_{F}%
	\text{\textrm{cos}}\alpha _{h}}\frac{\left( \tau _{0}-\tau _{z}\right) }{1-r_{R}^{b}r_{L}^{b}}%
\text{,} \label{CeeTIF}
\end{equation}

By substituting the above expressions in (\ref{GreenZZ2}), the Green's function for ferromagnetic and normal ($M = 0$) regions is obtained 
\begin{equation}
\hat{g}(x,x^{\prime })=\frac{-i}{\hbar v_{F}}\left( 
\begin{array}{cc}
\hat{g}_{ee}(x,x^{\prime }) & 0 \\ 
0 & \hat{g}_{hh}(x,x^{\prime })%
\end{array}%
\right) \text{,}  \label{Green_pelicula_TI}
\end{equation}%
where the Green's functions for electrons and holes sectors are given by
\begin{eqnarray}
\hat{g}_{ee/hh}(x,x^{\prime })=\left( \frac{N_{e/h}\mathrm{e}^{\pm i\left( x^{\prime }-x\right)
		k_{e/h}}}{2(1-r_{R}^{a/b}r_{L}^{a/b})\text{\textrm{cos}}\alpha _{e/h}}\right)\times\text{       } \label{Green_pelicula_eh}
\\
\left( 
\begin{array}{cc}
M_{+}^{2}IK & \mp siM_{+}M_{-}\mathrm{e}^{si\alpha }IL \\ 
\pm siM_{+}M_{-}\mathrm{e}^{-si\alpha }JK & M_{-}^{2}JL%
\end{array}%
\right) _{e/h}\notag\text{,}
\end{eqnarray}%
with the parameters
\begin{gather}
I_{e}=1+r_{L}^{a}\mathrm{e}^{s2ik_{e}x}\text{, \ } 
I_{h}=1-r_{L}^{b}\mathrm{e}^{-s2i\alpha _{h}}\mathrm{e}^{-s2ik_{h}x}\text{%
	,}  \notag \\
J_{e}=1-r_{L}^{a}\mathrm{e}^{s2i\alpha _{e}}\mathrm{e}^{s2ik_{e}x}\text{, \ }
J_{h}=1+r_{L}^{b}\mathrm{e}^{-s2ik_{h}x}\text{,}  \notag \\
K_{e}=1+r_{R}^{a}\mathrm{e}^{-s2ik_{e}x^{\prime }}\text{, \ }
K_{h}=1-r_{R}^{b}\mathrm{e}^{s2i\alpha _{h}}\mathrm{e}^{s2ik_{h}x^{\prime
}}\text{,}  \notag \\
L_{e}=1-r_{R}^{a}\mathrm{e}^{-s2i\alpha _{e}}\mathrm{e}%
^{-s2ik_{e}x^{\prime }}\text{, \ }
L_{h}=1+r_{R}^{b}\mathrm{e}^{s2ik_{h}x^{\prime }}\text{,}  \notag \\
N_{e/h}=(E_{F}\pm E)/{\sqrt{\left( E_{F}\pm E\right)^{2}-M^{2}}}\text{,}
\end{gather}
where $s=1$ for $x<x^{\prime }$, while $s=-1$ for $x^{\prime }<x$ and the subscripts are exchanged in the reflection coefficients of the previous expressions ($R\leftrightarrow L$). In the case of a semi-infinite  left(right) surface, the
Green's functions are obtained from (\ref{Green_pelicula_TI}) by making $%
r_{L}^{i}=0$ ($r_{R}^{i}=0$) since there are no reflection processes for open boundary conditions. 

\subsection{Superconducting regions\label{sec:app1B}}
Analogously, asymptotic solutions for the superconducting regions are calculated using the eigenvectors of the Hamiltonian (\ref{HBdG}) and the reflection coefficients associated with the boundary conditions (\ref{zzcond}). For the superconducting case and $M=0$ the Hamiltonian (\ref{HBdG}) has the spectrum
\begin{equation}
E=\pm \sqrt{\left( \hbar v_{F}\left\vert \mathbf{k}\right\vert -E_{F}\right)
	^{2}+\Delta _{0}^{2}}\text{,}
\end{equation}%
and eigenstates of the form
\begin{align}
\psi _{\varepsilon }^{qe}\left( x\right) & =\mathrm{e}^{\varepsilon
	ik_{qe}x}\left( u_{0}\hat{\varphi}_{\varepsilon }^{qe},-iv_{0}\sigma _{y}\hat{%
	\varphi}_{\varepsilon }^{qe}\right) ^{T}\text{,} \\
\psi _{\varepsilon }^{qh}\left( x\right) & =\mathrm{e}^{\varepsilon
	ik_{qh}x}\left( -iv_{0}\sigma _{y}\hat{\varphi}_{\varepsilon }^{qh},u_{0}\hat{%
	\varphi}_{\varepsilon }^{qh}\right) ^{T}\text{,}  \notag
\end{align}%
where the coherence factors are given by
\begin{eqnarray}
u_{0}&=&\sqrt{\frac{1}{2}\left( 1+\frac{\Omega }{E}\right) }\text{,}\,\,v_{0}=%
\sqrt{\frac{1}{2}\left( 1-\frac{\Omega }{E}\right) }\text{,}\\
\Omega&=&\sqrt{%
	E^{2}-\left\vert \Delta \right\vert ^{2}}\text{,}
\end{eqnarray}%
and the spinors $\hat{\varphi}_{\varepsilon }^{\mu }$ ($\mu =qe,qh$) by the expressions
 \begin{eqnarray}
 \hat{\varphi}_{\varepsilon }^{qe} &=&\left( 1,-\varepsilon i%
 \mathrm{e}^{\varepsilon i\alpha _{qe}}\right) ^{T}/\sqrt{2}\text{,}
 \label{espinor3} \\
 \hat{\varphi}_{\varepsilon }^{qh} &=&\left( \varepsilon i\mathrm{e}%
 ^{\varepsilon i\alpha _{qh}},1\right) ^{T}/\sqrt{2}\text{,}
 \label{espinor4} \\
 \mathrm{e}^{i\alpha _{\mu }} &\equiv &\hbar v_{F}\left( k_{\mu }+iq\right)
 /\left( E_{F}\pm E\,\right) \text{,}
 \end{eqnarray}%
with wave number in $x$
\begin{equation}
k_{qe/qh}=\text{\textrm{sgn}}\left( E_{F}\pm \Omega \right) \sqrt{\frac{\left(
		E_{F}\pm \Omega \right) ^{2}}{\hbar ^{2}v_{F}^{2}}-q^{2}}\text{.}
\end{equation}
In this case the reflection coefficients in (\ref{asymp}) for the boundary conditions $%
\psi _{L}^{\mu }\left( x_{L}\right) |_{\uparrow }=\psi _{R}^{\mu }\left(
x_{R}\right) |_{\downarrow }=0$ are ($\Gamma _{0}=v_{0}/u_{0}$)
\begin{eqnarray}
r_{L}^{a/b} &=&\mp \frac{\mathrm{e}^{\mp i\alpha _{qh}}-\Gamma _{0}^{2}%
	\mathrm{e}^{\mp i\alpha _{qe}}}{\mathrm{e}^{-i\alpha _{qh}}+\Gamma _{0}^{2}%
	\mathrm{e}^{i\alpha _{qe}}}\mathrm{e}^{\mp 2ik_{qe/qh}x_{L}}\text{,} \\
r_{R}^{a/b} &=&\pm \frac{\mathrm{e}^{\pm i\alpha _{qe}}-\Gamma _{0}^{2}%
	\mathrm{e}^{\pm i\alpha _{qh}}}{\mathrm{e}^{-i\alpha _{qe}}+\Gamma _{0}^{2}%
	\mathrm{e}^{i\alpha _{qh}}}\mathrm{e}^{\pm 2ik_{qe/qh}x_{R}}\text{,}  \notag \\
r_{L}^{c/d} &=&-\frac{2\Gamma _{0}\text{\textrm{cos}}\alpha _{qe/qh}}{%
	\mathrm{e}^{-i\alpha _{qh}}+\Gamma _{0}^{2}\mathrm{e}^{i\alpha _{qe}}}\mathrm{e%
}^{i\left( k_{qh}-k_{qe}\right) x_{L}}\text{,}  \notag \\
r_{R}^{c/d} &=&-\frac{2\Gamma _{0}\text{\textrm{cos}}\alpha _{qe/qh}}{%
	\mathrm{e}^{-i\alpha _{qe}}+\Gamma _{0}^{2}\mathrm{e}^{i\alpha _{qh}}}\mathrm{e%
}^{i\left( k_{qe}-k_{qh}\right) x_{R}}\text{,}  \notag
\end{eqnarray}%
and for the alternate boundary conditions
$\psi _{L}^{\mu }\left(
x_{L}\right) |_{\downarrow }=\psi _{R}^{\mu }\left( x_{R}\right) |_{\uparrow
}=0$
\begin{eqnarray}
r_{L}^{a/b} &=&\pm \frac{\mathrm{e}^{\mp i\alpha _{qe}}-\Gamma _{0}^{2}%
	\mathrm{e}^{\mp i\alpha _{qh}}}{\mathrm{e}^{i\alpha _{qe}}+\Gamma _{0}^{2}%
	\mathrm{e}^{-i\alpha _{qh}}}\mathrm{e}^{\mp 2ik_{qe/qh}x_{L}}\text{,} \\
r_{R}^{a/b} &=&\mp \frac{\mathrm{e}^{\pm i\alpha _{qh}}-\Gamma _{0}^{2}%
	\mathrm{e}^{\pm i\alpha _{qe}}}{\mathrm{e}^{i\alpha _{qh}}+\Gamma _{0}^{2}%
	\mathrm{e}^{-i\alpha _{qe}}}\mathrm{e}^{\pm 2ik_{qe/qh}x_{R}}\text{,}  \notag \\
r_{L}^{c/d} &=&-\frac{2\Gamma _{0}\text{\textrm{cos}}\alpha _{qe/qh}}{%
	\mathrm{e}^{i\alpha _{qe}}+\Gamma _{0}^{2}\mathrm{e}^{-i\alpha _{qh}}}\mathrm{e%
}^{i\left( k_{qh}-k_{qe}\right) x_{L}}\text{,}  \notag \\
r_{R}^{c/d} &=&-\frac{2\Gamma _{0}\text{\textrm{cos}}\alpha _{qe/qh}}{%
	\mathrm{e}^{i\alpha _{qh}}+\Gamma _{0}^{2}\mathrm{e}^{-i\alpha _{qe}}}\mathrm{e%
}^{i\left( k_{qe}-k_{qh}\right) x_{R}}\text{.}  \notag
\end{eqnarray}

The Green's functions of the superconducting system are given by the general expression (\ref%
{GreenZZ2}), and the constraint condition (\ref{LigaduraTI}) leads to the following relations
for the coefficient matrices
\begin{eqnarray}
\hat{C}_{\mu \nu } &=&\hat{C}_{\mu \nu }^{\prime }\text{, \ \ \ }\hat{C}%
_{hh}=X\hat{C}_{ee}\text{,}  \notag \\
\hat{C}_{eh} &=&Y\hat{C}_{ee}\text{, \ }\hat{C}_{he}=Z\hat{C}_{ee}\text{,}
\end{eqnarray}%
where the proportionality factors $X,Y$ and $Z$ depend only on the reflection coefficients
\begin{eqnarray}
X &=&\frac{\left( r_{L}^{c}r_{R}^{b }+r_{L}^{a}r_{R}^{c}\right) \left(
	r_{L}^{d }r_{R}^{c}+r_{L}^{a}r_{R}^{a}-1\right) }{\left(
	r_{L}^{a}r_{R}^{d }+r_{L}^{d }r_{R}^{b }\right) \left(
	r_{L}^{d }r_{R}^{c}+r_{L}^{b }r_{R}^{b }-1\right) }\text{,}
\\
Y &=&-\frac{r_{R}^{d }r_{L}^{c}+r_{L}^{d }r_{R}^{c}\left(
	r_{L}^{a}r_{R}^{a}+r_{L}^{b }r_{R}^{b }+r_{L}^{d
	}r_{R}^{c}-1\right) }{\left( r_{L}^{a}r_{R}^{d }+r_{L}^{d
	}r_{R}^{b }\right) \left( r_{L}^{d }r_{R}^{c}+r_{L}^{b
	}r_{R}^{b }-1\right) }  \notag \\
&&-\frac{r_{R}^{a}r_{R}^{b }r_{L}^{d
	}r_{L}^{c}+r_{L}^{a}r_{L}^{b }r_{R}^{d }r_{R}^{c}-r_{L}^{d
	}r_{L}^{c}r_{R}^{d }r_{R}^{c}}{\left( r_{L}^{a}r_{R}^{d
	}+r_{L}^{d }r_{R}^{b }\right) \left( r_{L}^{d
	}r_{R}^{c}+r_{L}^{b }r_{R}^{b }-1\right) }\text{,}  \notag \\
Z &=&\frac{r_{L}^{a}r_{R}^{c}+r_{L}^{c}r_{R}^{b }}{1-r_{L}^{d
	}r_{R}^{c}-r_{L}^{b }r_{R}^{b }}\text{,}  \notag
\end{eqnarray}
and the matrix $\hat{C}_{ee}$ is given by the expression
\begin{equation}
\hat{C}_{ee}=-\frac{2i}{\hbar v_{F}u_{0}^{2}}\frac{1}{Q^{2}-PR}\left( 
\begin{array}{cccc}
R & 0 & 0 & Q \\ 
0 & R & -Q & 0 \\ 
0 & Q & -P & 0 \\ 
-Q & 0 & 0 & -P%
\end{array}%
\right) \text{,}\label{CeeTI}
\end{equation}%
with the parameters
\begin{eqnarray}
P &=&A+XD+YG+ZJ\text{,} \\
Q &=&B+XE+YH+ZK\text{,}  \notag \\
R &=&C+XF+YI+ZL\text{,}  \notag \\
A &=&\left( 1-r_{L}^{a}r_{R}^{a}\right) \text{\textrm{cos}}\alpha
_{qe}+\Gamma _{0}^{2}r_{L}^{c}r_{R}^{c}\text{\textrm{cos}}\alpha _{qh}\text{,}
\\
B &=&-\Gamma _{0}\left( \left( 1-r_{L}^{a}r_{R}^{a}\right) \text{\textrm{cos}%
}\alpha _{qe}+r_{L}^{c}r_{R}^{c}\text{\textrm{cos}}\alpha _{qh}\right) \text{,}\notag
\\
C &=&\Gamma _{0}^{2}\left( 1-r_{L}^{a}r_{R}^{a}\right) \text{\textrm{cos}}%
\alpha _{qe}+r_{L}^{c}r_{R}^{c}\text{\textrm{cos}}\alpha _{qh}\text{,}\notag \\
D &=&-\Gamma _{0}^{2}\left( 1-r_{L}^{b}r_{R}^{b}\right) \text{%
	\textrm{cos}}\alpha _{qh}-r_{L}^{d}r_{R}^{d}\text{\textrm{cos}}%
\alpha _{qe}\text{,}\notag \\
E &=&\Gamma _{0}\left( \left( 1-r_{L}^{b}r_{R}^{b}\right) 
\text{\textrm{cos}}\alpha _{qh}+r_{L}^{d}r_{R}^{d}\text{\textrm{%
		cos}}\alpha _{qe}\right) \text{,}\notag \\
F &=&-\left( 1-r_{L}^{b}r_{R}^{b}\right) \text{\textrm{cos}}%
\alpha _{qh}-\Gamma _{0}^{2}r_{L}^{d}r_{R}^{d}\text{\textrm{cos}%
}\alpha _{qe}\text{,}\notag \\
G &=&\Gamma _{0}^{2}r_{L}^{c}r_{R}^{b}\text{\textrm{cos}}\alpha
_{qh}-r_{L}^{a}r_{R}^{d}\text{\textrm{cos}}\alpha _{qe}\text{,}\notag \\
H &=&-\Gamma _{0}\left( r_{L}^{c}r_{R}^{b}\text{\textrm{cos}}\alpha
_{qh}-r_{L}^{a}r_{R}^{d}\text{\textrm{cos}}\alpha _{qe}\right) \text{,}\notag
\\
I &=&r_{L}^{c}r_{R}^{b}\text{\textrm{cos}}\alpha _{qh}-\Gamma
_{0}^{2}r_{L}^{a}r_{R}^{d}\text{\textrm{cos}}\alpha _{qe}\text{,}\notag \\
J &=&\Gamma _{0}^{2}r_{L}^{b}r_{R}^{c}\text{\textrm{cos}}\alpha
_{qh}-r_{L}^{d}r_{R}^{a}\text{\textrm{cos}}\alpha _{qe}\text{,}\notag \\
K &=&-\Gamma _{0}\left( r_{L}^{b}r_{R}^{c}\text{\textrm{cos}}\alpha
_{qh}-r_{L}^{d}r_{R}^{a}\text{\textrm{cos}}\alpha _{qe}\right) \text{,}\notag
\\
L &=&r_{L}^{b}r_{R}^{c}\text{\textrm{cos}}\alpha _{qh}-\Gamma
_{0}^{2}r_{L}^{d}r_{R}^{a}\text{\textrm{cos}}\alpha _{qe}\text{.}\notag
\end{eqnarray}

In this case, the Green's function (\ref{GreenZZ2}) would be too extensive to write it explicitly. Again, in the case of a left (right) semi-infinite surface, the Green's functions are obtained from (\ref{GreenZZ2}) by making $%
r_{L}^{i}=0$ ($r_{R}^{i}=0$). 

\section{Dyson equation for coupling adjacent regions: Green's functions for FF and FS junctions\label{sec:app1C}}
The equilibrium Green's functions of a system, evaluated at the interface $x_0$ between two coupled regions (Fig.\ref{fig:regiones}) is obtained by the Dyson equation (\ref{eq:Dyson_equation}), which can be written in the simple form \cite{Gomez_2011,Casas_2019}
\begin{eqnarray}
\hat{G}_{L/R} &=&\hat{g}_{L/R}+\hat{g}(x_{L/R},x_{0}\mp \varepsilon ^{-})\hat{\Sigma}_{LR/RL}\times  \label{DY1} \\
&&\hat{M}%
_{R/L}g_{R/L,0}\hat{\Sigma}_{RL/LR}\hat{g}(x_{0}\mp \varepsilon
^{+},x_{L/R}^{\prime })\text{,}  \notag \\
\hat{G}_{LR/RL} &=&\hat{g}(x_{L/R},x_{0}\mp \varepsilon ^{-})\hat{\Sigma}_{LR/RL}\times \label{DY2}
\\
&&\hat{M}_{R/L}\hat{g}(x_{0}\pm \varepsilon ^{+},x_{R/L}^{\prime })%
\text{,}  \notag  \\
\hat{M}_{L/R} &=&\left[ 1-\hat{g}_{L/R,0}\hat{\Sigma}_{LR/RL}\hat{g}_{R/L,0}%
\hat{\Sigma}_{RL/LR}\right] ^{-1}\text{,}\notag
\end{eqnarray}%
where $\varepsilon ^{\pm}$ are infinitesimal scalars such that $0<\varepsilon ^{-}<\varepsilon ^{+}\ll 1$,  and were was used the following abbreviated notation: $\hat{G}_{L/R}=\hat{G}\left( x_{L/R},x_{L/R}^{\prime }\right) $, $\hat{G}%
_{LR/RL}=\hat{G}\left( x_{L/R},x_{R/L}^{\prime }\right) $, $\hat{g}_{L/R}=%
\hat{g}\left( x_{L/R},x_{L/R}^{\prime }\right) $, $\hat{g}_{L/R,0}=\hat{g}%
(x_{0}\mp \varepsilon ^{+},x_{0}\mp \varepsilon ^{-})$. 

As a first example, consider the case of the FF junction. The Green's functions of the decoupled regions $\hat{g}_{L/R}$ around $x_0=0$ can be obtained from (\ref{Green_pelicula_TI}) by making $r_{L/R}^{i}=0$ for left/right region. Hence, the Green's functions of the e/h sectors (\ref{Green_pelicula_eh}) for the left region take the form
\begin{eqnarray}
\hat{g}_{ee/hh}(-\varepsilon ^{+},-\varepsilon ^{-}) &=&\left( 
\begin{array}{cc}
0 & \pm 1 \\ 
0 & iL_{L}\mathrm{e}^{-i\alpha _{L}}%
\end{array}%
\right) _{e/h}\text{,}  \label{GM01} \\
\hat{g}_{ee/hh}(-\varepsilon ^{-},-\varepsilon ^{+}) &=&\left( 
\begin{array}{cc}
0 & 0 \\ 
\pm 1 & iL_{L}\mathrm{e}^{-i\alpha _{L}}%
\end{array}%
\right) _{e/h}\text{,}  \label{GM02} 
\end{eqnarray}
and for the right region
\begin{eqnarray}
\hat{g}_{ee/hh}(\varepsilon ^{+},\varepsilon ^{-}) &=&\left( 
\begin{array}{cc}
iL_{R}\mathrm{e}^{-i\alpha_{R} } & 0 \\ 
\pm 1 & 0%
\end{array}%
\right) _{e/h}\text{,}  \label{GM03}
\end{eqnarray}
with the factors
\begin{eqnarray}
L_{L,e/h}&=&\frac{E_{FL}\pm E-M_{L}}{\sqrt{\left( E\pm E_{FL}\right)
		^{2}-M_{L}^{2}}}\text{, \ }\\
L_{R,e/h}&=&\frac{E_{FR}\pm E+M_{R}}{\sqrt{\left(
		E\pm E_{FR}\right) ^{2}-M_{R}^{2}}}\text{.}
\end{eqnarray}

Thus, the Green's function of the coupled system given by (\ref{DY1}) takes the form
\begin{gather}
\hat{G}(-\varepsilon ^{+},-\varepsilon ^{-}) =\frac{-i}{\hbar v_{F}}\left( 
\begin{array}{cc}
\hat{G}_{ee} & 0 \\ 
0 & \hat{G}_{hh}%
\end{array}%
\right) \text{,} \\
\hat{G}_{ee/hh}=\frac{1}{D_{e/h}}\left( 
\begin{array}{cc}
L_{R}e^{-i\alpha _{R}} & \mp i \\ 
\pm iL_{L}L_{R}e^{-i\alpha _{L}}e^{-i\alpha _{R}} & L_{I}e^{-i\alpha _{L}}%
\end{array}%
\right) _{e/h}\text{,}  \notag \\
D_{e/h}=\left( 1+L_{L}L_{R}e^{-i\alpha _{L}}e^{-i\alpha _{R}}\right)
_{e/h}\text{,}  \notag
\end{gather}
where the roots of the denominator $D_{e/h}$ give rise to the dispersion relation of the QAHE  chiral state for a domain wall configuration. 

For the case of the FS junction, the Green's function at the right of the interface $\hat{g}_{R}(\varepsilon ^{+},\varepsilon ^{-})$ is replaced by the Green's function of the superconducting case with $r_{R}^{i}=0$
\begin{eqnarray}
&\hat{g}_R(\varepsilon ^{+},\varepsilon ^{-})=\label{superR}\\ &\frac{-1}{\hbar v_{F}}\left( 
\begin{array}{cccc}
\frac{i\left( 1-\Gamma _{0}^{2}\right) }{e^{i\alpha _{qe}}+\Gamma
	_{0}^{2}e^{-i\alpha _{qh}}} & 0 & -\frac{\Gamma _{0}\left( e^{i\alpha
		_{qe}}+e^{-i\alpha _{qh}}\right) }{e^{i\alpha _{qe}}+\Gamma _{0}^{2}e^{-i\alpha
		_{qh}}} & 0 \\ 
1 & 0 & 0 & 0 \\ 
-\frac{\Gamma _{0}\left( e^{i\alpha _{qe}}+e^{-i\alpha _{qh}}\right) }{%
	e^{i\alpha _{qe}}+\Gamma _{0}^{2}e^{-i\alpha _{qh}}} & 0 & \frac{i\left(
	1-\Gamma _{0}^{2}\right) }{e^{i\alpha _{qh}}+\Gamma _{0}^{2}e^{-i\alpha _{qe}}}
& 0 \\ 
0 & 0 & -1 & 0%
\end{array}%
\right) \notag \text{.} 
\end{eqnarray}

For the subsequent analytical calculation, the high doping limit for the superconducting region will be assumed ($e^{i\alpha _{\mu }}\sim 1$%
). Then, the Green's function of the coupled system takes the form
\begin{equation}
\hat{G}(-\varepsilon ^{+},-\varepsilon ^{-})=\frac{-i}{\hbar v_{F}D}\left( 
\begin{array}{cc}
\hat{g}_{ee} & \hat{g}_{eh} \\ 
\hat{g}_{eh}^{T} & \hat{g}_{hh}%
\end{array}%
\right) \text{,}  \label{GNS}
\end{equation}%
where Green's functions of the Nambu sectors are given by (subscripts denoting quasiparticles)
\begin{gather}
\hat{g}_{ee/hh} =\left( 
\begin{array}{cc}
X_{e/h} & \mp iY_{e/h} \\ 
\pm iL_{L,e/h}e^{-i\alpha _{e/h}}X_{e/h} & L_{L,e/h}e^{-i\alpha
	_{e/h}}Y_{e/h}%
\end{array}%
\right) \text{,} \\
\hat{g}_{eh} =\Delta \left( 
\begin{array}{cc}
i & L_{L,h}e^{-i\alpha _{h/e}} \\ 
-L_{L,e}e^{-i\alpha _{e/h}} & iL_{L,e}L_{L,h}e^{-i\alpha _{e}}e^{-i\alpha
	_{h}}%
\end{array}%
\right) \text{,} \\
X_{e/h} =\left( \Omega +L_{L,h/e}Ee^{-i\alpha _{h/e}}\right) ,\\
Y_{e/h}=\left( E+L_{L,h/e}e^{-i\alpha _{h/e}}\Omega \right)\text{,}
\end{gather}
and the denominator $D$ by
\begin{eqnarray}
D&=&E\left( L_{L,e}L_{L,h}e^{-i\alpha_e }e^{-i\alpha_h }+1\right) \\
 &&+\Omega \left(
L_{L,e}e^{-i\alpha_e }+L_{L,h}e^{-i\alpha_h }\right) \notag \text{,}
\end{eqnarray}
which leads to the dispersion relation of the IBS states \cite{Herrera_2009,Casas_2019_1,Casas_2019}
\begin{eqnarray}
E_{\text{IBS}} &=&\pm \frac{\left\vert \Delta \right\vert }{\sqrt{1-C^{2}}}%
\text{,}\label{IBSTI} \\
C &=&\frac{L_{L,e}L_{L,h}e^{-i\alpha _{e}}e^{-i\alpha _{h}}+1}{%
	L_{L,e}e^{-i\alpha _{e}}+L_{L,h}e^{-i\alpha _{h}}}\text{.}
\end{eqnarray}

\begin{figure}[h]
	\centering
	\includegraphics[width=1\columnwidth]{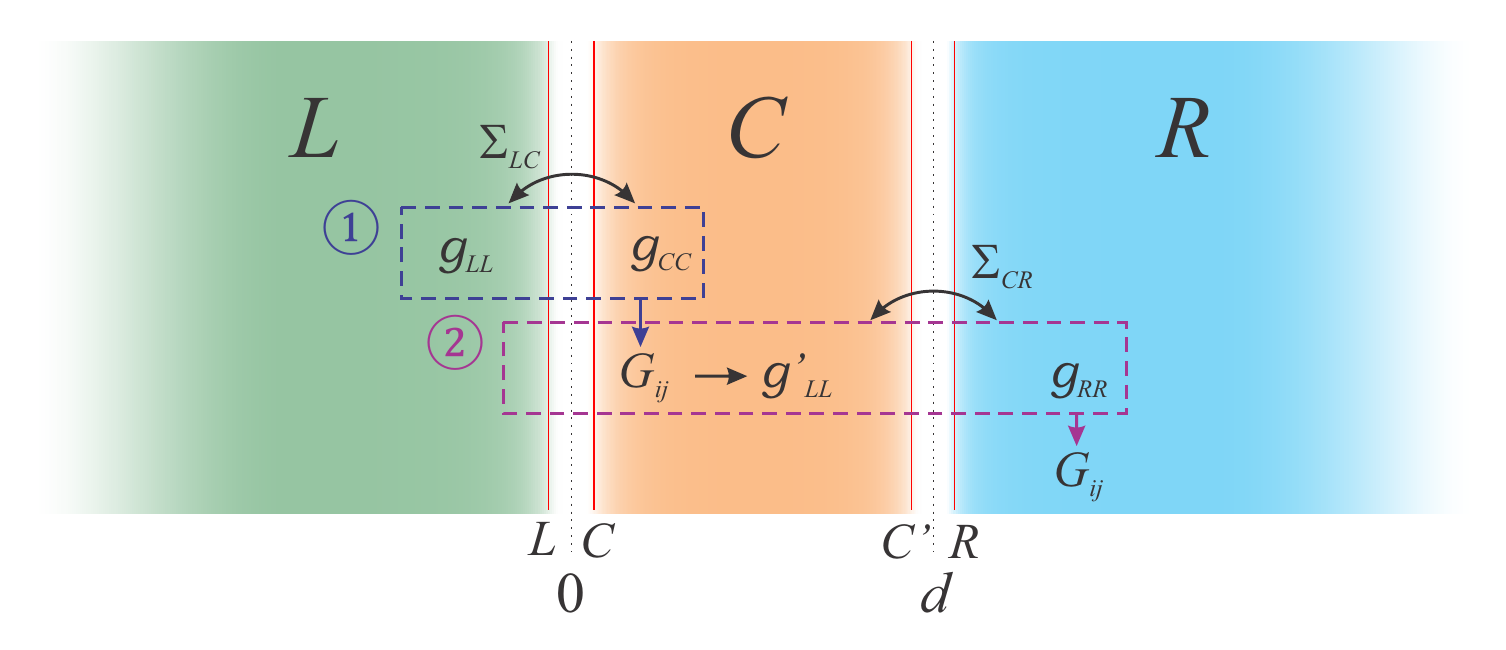}
	\caption{Schematic representation of the equilibrium Green's functions calculation for a junction with a finite central region and two semi-infnite lateral leads. First, the Green's functions of the $L$ and $C$ regions are coupled with the Dyson equation. Then, the resulting Green's functions are taken as the new left-input for a second Dyson equation to finally obtain the Green's functions of the complete system. } \label{fig:regiones2}
\end{figure}

This reduces to the expression (\ref{ChiralMajorana}) for zero doping levels of the two regions \cite{Casas_2019}. For the case of junctions with more than two regions, the Dyson equation can be implemented sequentially to each interface to obtain the equilibrium Green's functions of the entire system. Figure \ref{fig:regiones2} illustrates the coupling process for the case of a junction of three regions. For the first interface at $x=0$, the equilibrium Green's functions of the left ($L$) and central ($C$) regions are the ``left'' and ``right'' inputs in equations (\ref{DY1}-\ref{DY2}) to obtain the coupled Green's functions of the $L$-$C$ subsystem. Following the same procedure for the second interface at $x=d$, the Green's functions of the $L$-$C$ subsystem and those of the right region ($R$) are the new ``left'' and ``right'' inputs in the same equations, obtaining the total equilibrium Green's functions of the junction.

\section{Differential conductance derivation \label{sec:app2}}
For a system consisting of three regions as showed in Fig. \ref{fig:regiones2}, the Hamiltonian has the form \cite{Yeyati_1996,Casas_2019_1,Casas_2019} 
\begin{equation}
H=H_{L}+H_{C}+H_{R}+H_{T_{L}}+H_{T_{R}}\text{,}\label{AHTI}
\end{equation}%
with $H_{L,C,R}$ the Hamiltonians of the three decoupled regions
(left ($L$), central ($C$) and right ($R$)) and $H_{T_{L,R}}$ the tunneling Hamiltonians \ref{eq:hopping} for the left and right interfaces
\begin{align}
H_{T_{L}}(\tau )& =t\int \mathrm{d}q\mathrm{e}^{i\phi _{L}(\tau )/2} 
\hat{c}_{q,L\downarrow }^{\dagger }\hat{b}_{q,L\uparrow } +\text{h.c.,}
\label{eq:hop3} \\
H_{T_{R}}(\tau )& =t\int \mathrm{d}q\mathrm{e}^{i\phi _{R}(\tau
	)/2} \hat{c}_{q,R\uparrow }^{\dagger }\hat{b}_{q,R\downarrow } +\text{h.c.,}
\label{eq:hop4}
\end{align}%
where $t=\hbar v_{F}$, $\phi _{L(R)}(\tau )=\phi _{0}+2(\mu _{L(R)}-\mu
_{C})\tau /\hbar $ the gauge phases induced by the gradient of the
chemical potential, the $\hat{c}_{q,\nu \sigma }$, with $\nu =L,R$ and $%
\sigma =\uparrow ,\downarrow $, are the annihilation operators for electrons at the edges of the left and right regions with wave number $%
q$, and the $\hat{b}_{q,\nu \sigma }$ are the annihilation operators at the edges of the central region. In the Heisenberg picture, the average current at the left interface is given by
\begin{gather}
I(\tau )=-e\left\langle \frac{d}{d\tau }N_{L}(\tau )\right\rangle \\
=it\frac{e}{\hbar }\int \mathrm{d}q\left[ \left\langle \hat{c}%
_{q,L\downarrow }^{\dag }(\tau )\hat{b}_{q,L\uparrow }(\tau )\right\rangle
-\left\langle \hat{b}_{q,L\uparrow }^{\dagger }(\tau )\hat{c}_{q,L\downarrow
}(\tau )\right\rangle \right] \text{,}  \notag
\end{gather}%
which can be expressed in terms of Keldysh Green's functions as
\begin{equation}
\hat{G}_{q,ij}^{\alpha \beta }\left( \tau _{\alpha },\tau _{\beta }^{\prime
}\right) =-i\left\langle T_{c}\left[ \hat{D}_{q,i}\left( \tau _{\alpha
}\right) \hat{D}_{q,j}^{\dag }\left( \tau _{\beta }^{\prime }\right) \right]
\right\rangle \text{,}
\end{equation}%
where $i,j=L,C,C^{\prime },R$ are the border indexes of each region, $\alpha ,\beta $ indicate the Keldysh temporal branches, $T_{c}$
is the Keldysh time-ordering operator and where the following vector operators were defined  
\begin{eqnarray}
\hat{D}_{q,i}\left( \tau \right) &=&\left( \hat{d}_{q,i\uparrow }\left( \tau
\right) ,\hat{d}_{q,i\downarrow }\left( \tau \right) ,\hat{d}_{q,i\uparrow
}^{\dag }\left( \tau \right) ,\hat{d}_{q,i\downarrow }^{\dag }\left( \tau
\right) \right) ^{T}, \\
\hat{D}_{q,i}^{\dag }\left( \tau \right) &=&\left( \hat{d}_{q,i\uparrow
}^{\dag }\left( \tau \right) ,\hat{d}_{q,i\downarrow }^{\dag }\left( \tau
\right) ,\hat{d}_{q,i\uparrow }\left( \tau \right) ,\hat{d}_{q,i\downarrow
}\left( \tau \right) \right) ,
\end{eqnarray}%
according to relations $\ \hat{d}_{q,L\sigma }(\tau )=\hat{c}_{q,L\sigma
}(\tau )$, $\hat{d}_{q,C\sigma }(\tau )=\hat{b}_{q,L\sigma }(\tau )$, $\hat{d%
}_{q,C^{\prime }\sigma }(\tau )=\hat{b}_{q,R\sigma }(\tau )$, $\hat{d}%
_{q,R\sigma }(\tau )=\hat{c}_{q,R\sigma }(\tau )$. Considering a stationary situation, the average current can be written in energy space as ($\hat{t}=\Sigma _{LR}$)
\begin{equation}
I=\frac{e}{2h}\int \mathrm{d}q\mathrm{d}E\mathrm{Tr}\left( \tau _{z}\left[ 
\hat{t}\hat{G}_{q,CL}^{+-}(E)-\hat{t}^{\dag }\hat{G}_{q,LC}^{+-}(E)%
\right] \right) \text{.}
\end{equation}

This expression can also be applied for a two-region system with a single interface, by considering the union of regions $R$ and $C$ as a new region $R$, as explained in the previous section for Green's functions (this equation coincides with Eq.(\ref{current}) by changing index $C$ by $R$). By using the following Dyson equations
\begin{eqnarray}
\hat{G}_{CL}^{+-}(E) &=&\hat{G}_{CC}^{+-}(E)\hat{t}^{T}\hat{g}%
_{L}^{a}(E)+\hat{G}_{CC}^{r}(E)\hat{t}^{T}\hat{g}_{L}^{+-}(E)\text{,}
\label{Dy3} \\
\hat{G}_{LC}^{+-}(E) &=&\hat{g}_{L}^{+-}(E)\hat{t}\hat{G}_{CC}^{a}(E)+%
\hat{g}_{L}^{r}(E)\hat{t}\hat{G}_{CC}^{+-}(E)\text{,}  \label{Dy4}
\end{eqnarray}%
the average current can be written in terms of local Green's functions as
\begin{eqnarray}
I&=&\frac{e}{2h}\int \mathrm{d}q\mathrm{d}E\mathrm{Tr}\left( \tau _{z}\hat{t}^{\dag }\left[ \hat{g}_{q,L}^{+-}\left( E\right) \hat{t}\hat{G}%
_{q,CC}^{-+}\left( E\right)\right.\right.\\ 
&&\left. \left.-\hat{g}_{q,L}^{-+}\left( E\right) \hat{t}\hat{G}_{q,CC}^{+-}\left( E\right) \right] \right) \text{,}\notag
\end{eqnarray}%
where the unperturbed Keldysh Green's functions are given by the relations 
\begin{eqnarray}
\hat{g}_{q,i}^{+-}\left( E\right) &=&2\pi i\hat{\rho}_{q,i}\left( E\right) 
\hat{n}_{i}\left( E\right) \text{,}  \label{Kmn} \\
\hat{g}_{q,i}^{-+}\left( E\right) &=&-2\pi i\hat{\rho}_{q,i}\left( E\right)
\left( \tau _{0}-\hat{n}_{i}\left( E\right) \right) \text{,}  \label{Knm}
\end{eqnarray}%
being $\hat{\rho}_{q,i}=\mp\mathrm{Im}(\hat{g}
_{q,i}^{r(a)})/\pi $ the DOS matrix of the $i$ electrode and $\hat{n}_{i}$ the quasiparticle occupation matrix
\begin{eqnarray}
\hat{n}_{i}(E) &=&\mathrm{diag}(n_{i,e}(E)\tilde{\sigma}_{0},n_{i,h}(E)%
\tilde{\sigma}_{0})\text{,}
\end{eqnarray}%
with $n_{i,e/h}\left( E\right) =\left[ 1+\mathrm{e}^{\left( E\pm \left( \mu
	_{i}-E_{Fi}\right) \right) /k_{B}T}\right] ^{-1}$ the Fermi-Dirac functions for electrons/holes (from here we will consider the limit $T\rightarrow 0$).
By means of the following Dyson equation ($\gamma =+-,-+$)
\begin{equation}
\hat{G}_{q,CC}^{\gamma }=\hat{G}_{q,CC}^{r}\hat{t}^{\dag }\hat{g}%
_{q,L}^{\gamma }\hat{t}\hat{G}_{q,CC}^{a}+\hat{G}_{q,CC^{\prime }}^{r}%
\hat{t}\hat{g}_{q,R}^{\gamma }\hat{t}^{\dag }\hat{G}_{q,C^{\prime
	}C}^{a}\text{,}  \label{DyK3}
\end{equation}
the expression for the electrical current takes the form
\begin{gather}
I=\frac{2\pi ^{2}e}{h}\int \mathrm{d}q\mathrm{d}E\mathrm{Tr}\left\{ \tau _{z}%
\hat{t}^{\dag }\hat{\rho}_{q,L}\right. \label{eq:IL}\\
\left( \left[ \hat{n}_{L}\left( \hat{t}\hat{G}_{q,CC^{\prime }}^{r}\hat{t}\hat{\rho}_{q,R}\right) -\left( \hat{t}\hat{G}_{q,CC^{\prime }}^{r}%
\hat{t}\hat{\rho}_{q,R}\right) \hat{n}_{R}\right] \hat{t}^{\dag }%
\hat{G}_{q,C^{\prime }C}^{rT}+\right.  \notag \\
\left. \left. \left[ \hat{n}_{L}\left( \hat{t}\hat{G}_{q,CC}^{r}\hat{t}^{\dag }\hat{\rho}_{q,L}\right) -\left( \hat{t}\hat{G}_{q,CC}^{r}%
\hat{t}^{\dag }\hat{\rho}_{q,L}\right) \hat{n}_{L}\right] \hat{t}%
\hat{G}_{q,CC}^{rT}\right) \right\} \text{.}  \notag
\end{gather}

This current incorporates the contribution of transport processes between states with spin  $\uparrow $ on the left and spin $\downarrow $ to the right of each contact ($\uparrow
_{L}\leftrightarrows $ $\downarrow _{R}$). However, under normal conditions on the surface of a TI, both spin projections symmetrically contribute to the transport. Then, it is necessary to consider the contribution of the opposite-spin processes ($%
\downarrow _{L}\leftrightarrows $ $\uparrow _{R}$), which introduces an additional factor of $2$. Again, the electric potential $V$ is introduced as a shift between the chemical potentials of the electrodes ($\mu _{L}=E_{FL}+eV$ y $\mu _{R}=E_{FR}$)\ and the Fermi levels of the regions are controlled independently through gates. The normalized differential conductance is given by 
\begin{eqnarray}
\sigma &=&\sigma _{Q}+\sigma _{A}\text{,}  \label{condTI} \\
\sigma _{Q}&=&\frac{4}{\sigma _{0}}\frac{e^{2}}{h}%
\int \mathrm{d}q\mathrm{Tr}\left[\bar{\rho}%
_{q,L}\hat{G}_{q,CC^{\prime }}^{r}\bar{\rho}%
_{q,R}\hat{G}_{q,C^{\prime }C}^{a}\right] \text{,} \notag\\
\sigma _{A}&=&\frac{4}{\sigma _{0}}\frac{e^{2}}{h}%
\int \mathrm{d}q\mathrm{Tr}\left(\bar{\rho}_{q,L}\tau _{z}\hat{G}_{q,CC}^{r}\bar{\rho}_{q,L}\hat{G}_{q,CC}^{a}\tau_{z}\right. \notag \\
&& \left. - \tau _{z}\bar{\rho}_{q,L}\hat{G}_{q,CC}^{r}\bar{\rho}_{q,L}\tau _{z}\hat{G}_{q,CC}^{a}\right) \text{,}  \notag
\end{eqnarray} 
and in terms the Nambu space components 
\begin{eqnarray}
\sigma _{Q} &=&\frac{4e^{2}}{h\sigma _{0}}\int\mathrm{d}q\mathrm{Tr}\left[ \mathrm{Re}\left\{ \bar{\rho}%
_{q,Lee} {}\right. \right.  \notag  \\
&&\left. \left. \left(\hat{G}_{q,CC^{\prime }ee}^{r}\left[ \bar{\rho}_{q,Ree}\hat{G}%
_{q,C^{\prime }Cee}^{a}-\bar{\rho}_{q,Reh}\hat{G}_{q,C^{\prime
	}Che}^{a}\right] \right. \right.\right.   \notag \\
&&-\left. \left. \left. \hat{G}_{q,CC^{\prime }eh}^{r}\left[ \bar{\rho}%
_{q,Rhe}\hat{G}_{q,C^{\prime }Cee}^{a}-\bar{\rho}_{q,Rhh}\hat{G}%
_{q,C^{\prime }Che}^{a}\right] \right) \right\} \right],  \notag\\
\sigma _{A}&=&\frac{8e^{2}}{h\sigma _{0}}\int\mathrm{d}q\mathrm{Tr}\left[ \mathrm{Re}\left\{ \bar{\rho%
}_{q,Lee}\hat{G}_{q,CCeh}^{r}\bar{\rho}_{q,Lhh}\hat{G}_{q,CChe}^{a}\right\} %
\right].\notag
\end{eqnarray}
Here, the following matrices were defined 
\begin{equation}
\bar{\rho}_{q,L}\equiv\pi \hat{t}^{\dagger }\hat{\rho}_{q,L}\hat{t}, \ \
\bar{\rho}_{q,R}\equiv\pi \hat{t}\hat{\rho}_{q,R}\hat{t}^{\dagger},
\end{equation}
and the two components have been normalized to ballistic conductance per unit of surface length of a TI, $\sigma
_{0}\left( V\right) =2e^{2}\left( E_{F}+eV\right) /h\pi \hbar v_{F}$. 
This expression for the conductance can be approximated to those of the case of two coupled semi-infinite regions, at the limit when the width of the central region tends to zero ($d\rightarrow 0$), and by making the parameters of this region equal to those of any of the electrodes. For example, for a normal left region and superconducting right region, the component $\sigma_{Q}$ includes terms that involve the DOS of both electrodes and is generally associated with electron-electron transport processes by direct transport ($\propto\hat{\rho}_{q,Lee}\hat{\rho}_{q,Ree}$) or by pair-creation/-annihilation process as intermediate state ($\propto \hat{\rho}_{q,Lee}\hat{\rho}_{q,Reh/he}$), in addition to electron-hole conversion processes in the superconductor ($\propto \hat{\rho}_{q,Lee}\hat{\rho}_{q,Rhh}$) by branch crossing. On the other hand, the product $\hat{\rho}_{q,Lee}\hat{\rho}_{q,Lhh}$ in the second component involves only the left electrode and is associated with electron-hole conversion processes such as Andreev reflections with a probability proportional to $| \hat{G}_{q,CChe}^{r}|^{2}$ \cite{Yeyati_1996}.

\bibliography{article}

\end{document}